\documentclass[onecolumn,amsmath,amssymb,prd,reprint,longbibliography,floatfix,8pt]{revtex4-1}

\usepackage{cancel}
\usepackage{enumitem}
\usepackage[utf8x]{inputenc}
\usepackage{graphicx}
\usepackage{dcolumn}
\usepackage{braket}
\usepackage{bm}
\usepackage[switch]{lineno}
\usepackage{float} 
\usepackage{subfigure}
\usepackage{tikz}
\usepackage{multirow}
\usetikzlibrary{arrows.meta}
\usetikzlibrary{arrows,decorations.pathmorphing,backgrounds,positioning,fit,petri}

\begin{document}

\title{Charting new regions of Cobalt's chemical space with maximally large magnetic anisotropy: A computational high-throughput study}

\author{Lorenzo A. Mariano}
\affiliation{School of Physics, AMBER and CRANN Institute, Trinity College, Dublin 2, Ireland}
\author{Vu Ha Anh Nguyen}
\affiliation{School of Physics, AMBER and CRANN Institute, Trinity College, Dublin 2, Ireland}
\author{Valerio Briganti}
\affiliation{School of Physics, AMBER and CRANN Institute, Trinity College, Dublin 2, Ireland}
\author{Alessandro Lunghi}
\email{lunghia@tcd.ie}
\affiliation{School of Physics, AMBER and CRANN Institute, Trinity College, Dublin 2, Ireland}

\begin{abstract}
{\bf Magnetic anisotropy slows down magnetic relaxation and plays a prominent role in
the design of permanent magnets.
Coordination compounds of Co(II) in particular exhibit large
magnetic anisotropy in the presence of low-coordination environments and have been
used as single-molecule magnet prototypes. However, only a limited sampling of
Cobalt’s vast chemical space has been performed, potentially obscuring alternative chemical routes toward large magnetic anisotropy. Here we perform a computational high-throughput exploration of Co(II)'s chemical space in search of new single-molecule magnets. We automatically assemble a diverse
set of $\sim$15,000 novel complexes of Co(II) and fully characterize them with multi-reference ab initio
methods. More than 100 compounds exhibit magnetic anisotropy comparable to 
or larger than leading known compounds. The analysis of these results shows that compounds with record-breaking magnetic anisotropy can also be achieved with coordination four or higher, going beyond the established paradigm of two-coordinated linear complexes.}
\end{abstract}

\maketitle
\section*{Introduction}

Magnetic anisotropy describes how the energy of a magnetic moment varies with its orientation in space and is at the origin of slow magnetic relaxation\cite{Lunghi2022} and permanent magnetism \cite{Coey2010-qj}. At the microscopic level, magnetic anisotropy arises from the interaction between the spin, $S$, and orbital motion, $L$, of an ion's unpaired electrons, known as spin-orbit coupling $\hat{H}_{SOC}=\lambda L \cdot S$. Once the symmetry of the space surrounding the ion is broken by embedding it in a solid-state lattice, spin-orbit coupling establishes a preferential direction in space for the ion's magnetic moment. Among the chemical compounds exhibiting this property, Cobalt represents the prime example among transition metals and its large magnetic anisotropy has been reported for materials ranging from elemental Co\cite{ST} and intermetallic alloys\cite{Mohapatra2020} to molecules\cite{Long2018} and surface-adsorbed atoms\cite{Gambardella2003}. 

In the case of molecules and atoms in particular, record values of anisotropy have been observed\cite{Long2018,Gambardella2003}. Such groundbreaking results had been achieved by maximizing the value of $L$, which correlates with spin-orbit coupling strength. Whilst most coordination geometries lead to a quench of an ion's orbital angular momentum, low coordination such as two-coordinated linear Co(II) ions can maintain the maximally allowed value of $L=3$ and therefore exhibit large magnetic anisotropy\cite{Long2018}. Despite these extensive research efforts, this design rule poses serious challenges. Indeed, synthesizing linear two-coordinate Co compounds is quite challenging and only four compounds have been reported to date\cite{Long2018,Yao2016}. Moreover, low coordination environments for Cobalt ions lead to very reactive compounds limiting the scope of these molecules outside the confines of fundamental science. Although there is ample scope for the exploration of other chemical strategies for the realization of highly anisotropic Cobalt-based molecules, the intrinsic rarity of magnetism\cite{Vedmedenko2020} and the vastness of the chemical space hamper further progress\cite{Reymond2015}. 

Computational approaches and large-scale ab initio screening frameworks in particular are becoming leading tools to accelerate the discovery of new materials with desired properties\cite{Curtarolo2013,Nandy2021,Ale2022,Duan2022}. Seminal attempts to use such high-throughput approaches in the field of magnetism have targeted the determination of the spin ground state of spin crossover compounds\cite{Janet2018} and solid-state Heusler-type magnets\cite{Sanvito2017}. However, these studies are invariably performed with Density Functional Theory (DFT), which is known to lead to qualitatively wrong predictions of magnetic anisotropy in coordination compounds\cite{Atanasov2015}. On the other hand, attempts to systematically explore different coordination geometries for Co(II) single-ion complexes with accurate multi-reference ab initio methods have been confined to model systems with a single ligand or a handful of structures from crystallography databases\cite{GomezCoca2013,Chopek2006,Lococciolo2024,Gupta2023,Moseley2022}. Despite being very informative for the molecules at hand, this approach cannot account for the vast chemical and structural diversity of organic ligands' chemical space. 

\begin{figure*}
\centering
\includegraphics[scale=0.85]{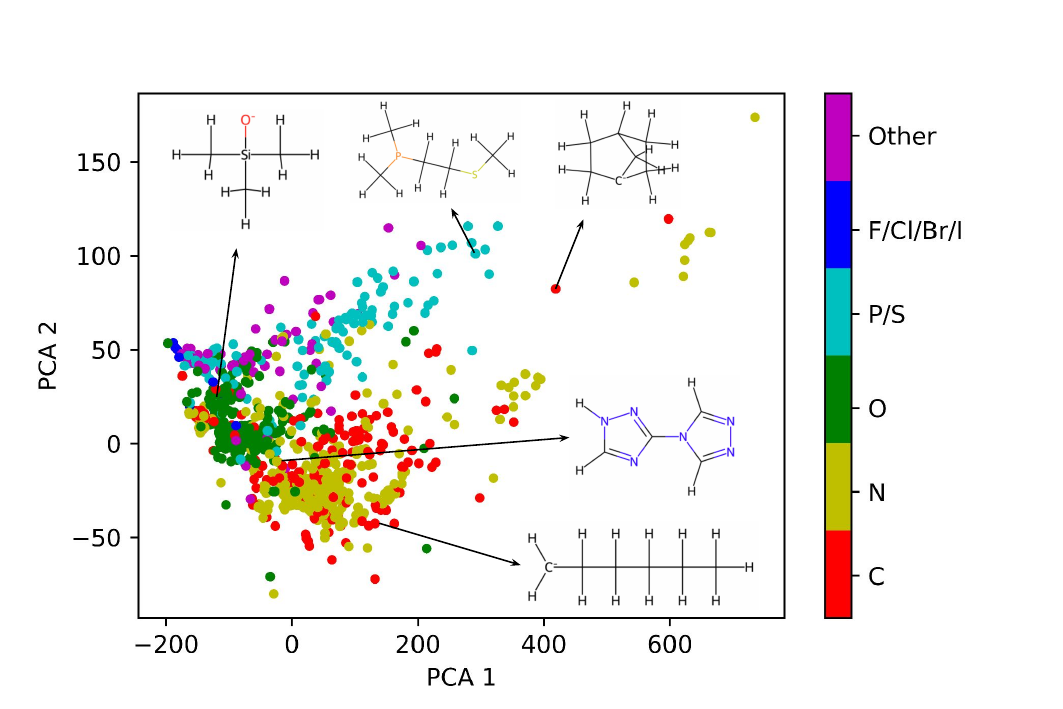}
\caption{Principal Component Analysis (PCA) for the whole set of monodentate ligands with less than 20 atoms. Different colours are associated with the different nature of the bonding atom as reported in the colourbar. The representation of a random selection of ligands is also reported.}
\label{PCA}
\end{figure*}
In this study, we remove the main roadblock to the systematic simulation of magnetic anisotropy by establishing a multi-reference ab initio high-throughput framework. Our study addresses around 15,000 Co(II) single-ion coordination compounds and individuates tens of candidates with record-large magnetic anisotropy. Most importantly, these record values are also achieved with coordination numbers higher than two and unprecedented coordination geometries. A new general design rule emerges from data with the potential of extending the scope of chemical synthesis in achieving molecules with simultaneous large magnetic anisotropy and high chemical stability.

\section*{Results}

\textbf{High-throughput ab initio screening.} The first step of our computational strategy is collecting all molecules containing Cobalt from the Cambridge Crystallographic Structural Database (CCSD)\cite{Groom2016}. We then select all crystals that only contain single-ion Co complexes and extract the associated organic ligands. The latter are then classified by the number of binding atoms, leading to a set of 1,423 mono-dentate ligands deposited in COMPASS\_lig2 database. The total number of coordination of compounds that can be generated with such a set of ligands is virtually infinite due to the combinatorial nature of the problem and constraints on the coordination motifs are introduced to make a systematic computational study possible. Fig. \ref{PCA} reports the results of a Principal Component Analysis (PCA) tailored to display the structural similarity between ligands. Based on this and capping the maximum size of the ligands to 20 atoms, the Furthest Point Sampling (FPS) method is used to select a subset of ligands with maximal chemical diversity, leading to a final set of 208 ligands that well represent the coordination chemistry of Co as available in CCSD. These ligands are deposited in COMPASS\_lig1 database.

Cobalt compounds with the largest magnetic anisotropy to date are found among linear two-coordinated and distorted tetrahedral ones. Based on this, we use the software MolSimplify\cite{molSimplify} to assemble a comprehensive dataset of tetrahedral Co(II)-based compounds with general brute formula CoA$_2$B$_2$, and linear two-coordinated compounds of the form Co(A)$_2$. Only two different ligands at the time are allowed for the tetrahedral geometry with the two-fold objective of making the final compounds more likely to be successfully synthesized and constraining the total number of compounds to be simulated. The ligands A and B are systematically chosen from our COMPASS\_lig1 dataset, leading to 21,528 four-coordinate compounds and 208 two-coordinate compounds. 

The total molecular charge is set based on the ligands' charges, determined with the help of DFT, and considering Cobalt in its 2+ oxidation state. All compounds were considered in their high-spin configuration, known for exhibiting large magnetic anisotropy in the chosen crystal field geometries\cite{KumarSahu2023}. The structure of each compound is optimized without imposing any constraint on the molecular symmetry. The final geometries and molecular orbitals are then used to run a multi-reference calculation and compute the spin Hamiltonian parameters describing the magnetic properties of the entire set (\textit{vide infra}). 

The total number of compounds which survive both geometry optimization and multi-reference calculation is 15,559 for coordination number four and 49 for coordination number two. The lower success rate for coordination two is due to the high instability of this low coordination number, often resulting in unacceptable optimized geometries. In the following, we will refer to the set obtained by combining 4 ligands as \textbf{Set-1} and the set obtained by combining 2 ligands as \textbf{Set-2}. The former constitutes our COMPASS\_set1 database, while the latter is stored in the COMPASS\_set2 database.\\

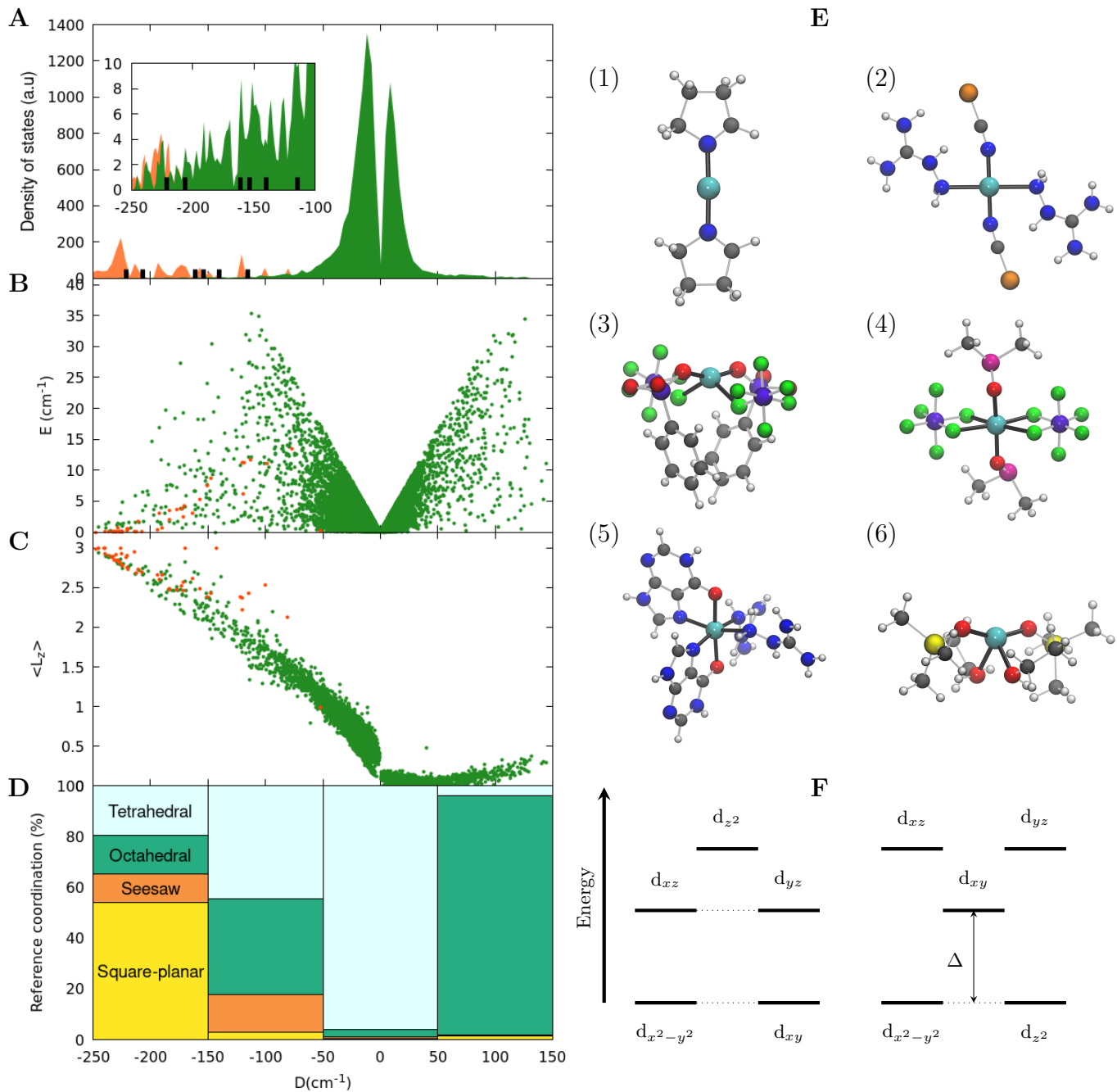
\begin{figure*}
 
\begin{tikzpicture}

\node[anchor=south west, inner sep=0] (image) at (-4,0.5) {\includegraphics[width=0.5\textwidth]{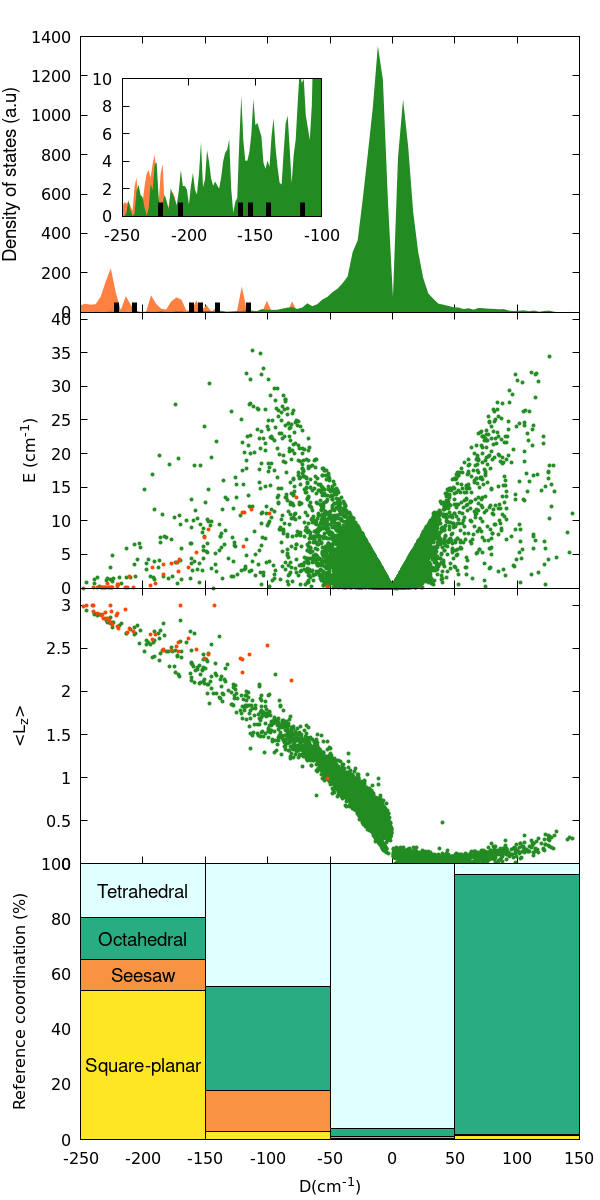}};

\node[anchor=south west, inner sep=0] (image) at (6.2,13.2)
{\includegraphics[width=0.11 \textwidth]{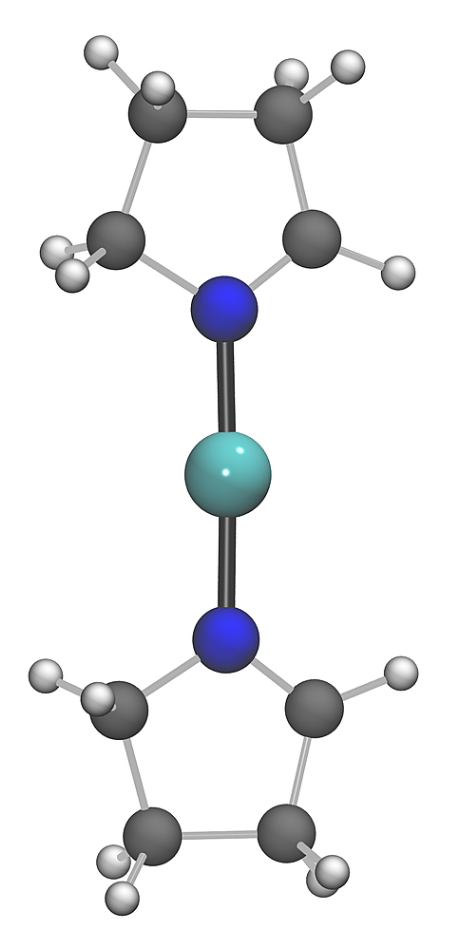}};
\node at (5.5,17) {\large{(1)}};

\node[anchor=south west, inner sep=0] (image) at (9.5,13.0)
{\includegraphics[width=0.25 \textwidth]{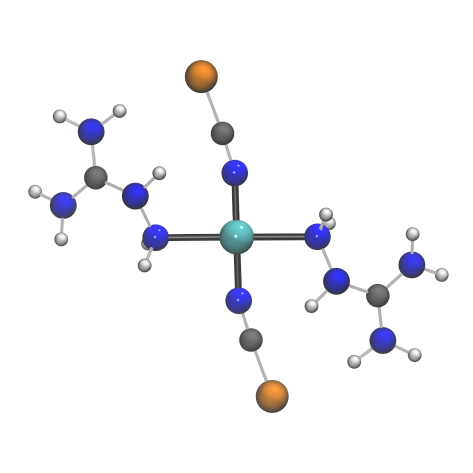}};
\node at (10.0,17) {\large{(2)}};

\node[anchor=south west, inner sep=0] (image) at (5.7,10.0)
{\includegraphics[trim={0 0 0 0cm},clip,width=0.17\textwidth]{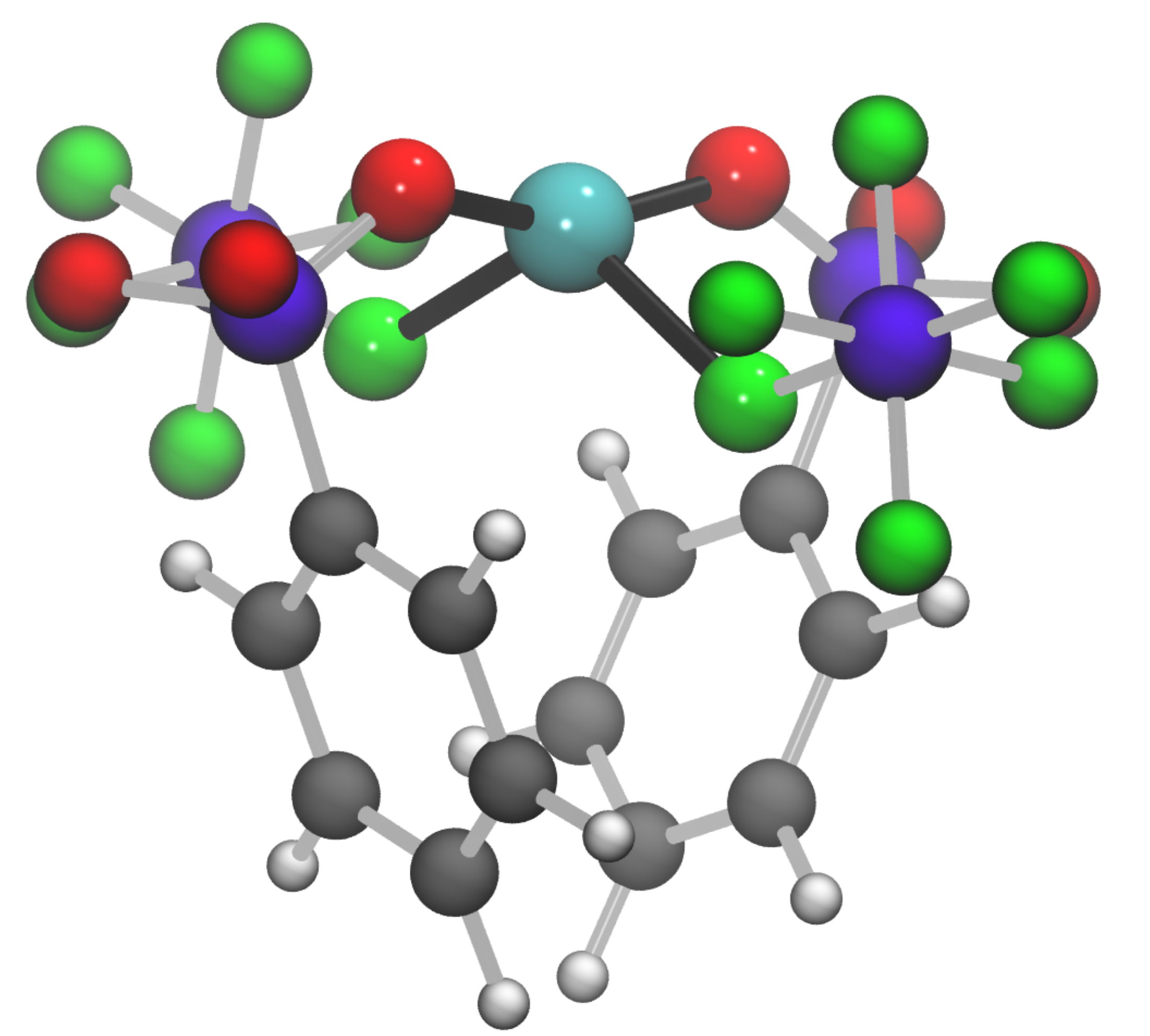}};
\node at (5.5,13.0) {\large{(3)}};

\node[anchor=south west, inner sep=0] (image) at (10.2,9.8)
{\includegraphics[width=0.18 \textwidth]{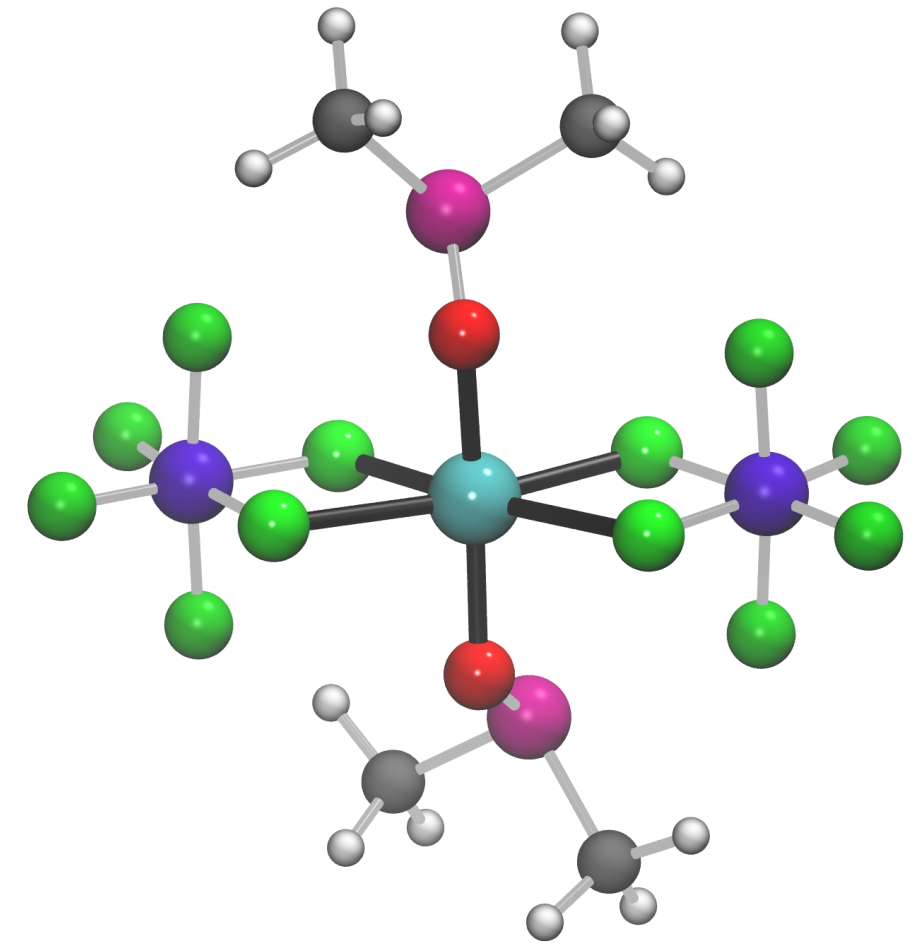}};
\node at (10,13.0) {\large{(4)}};

\node[anchor=south west, inner sep=0] (image) at (5.7,6.2)
{\includegraphics[width=0.19 \textwidth]{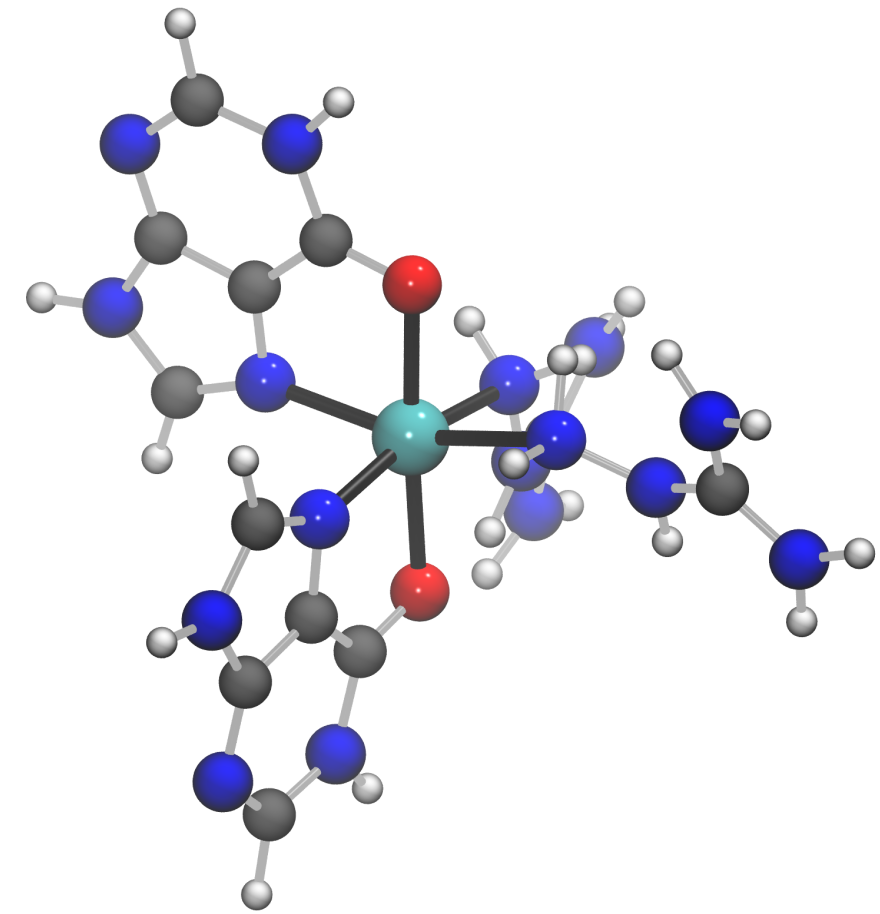}};
\node at (5.5,9.6) {\large{(5)}};

\node[anchor=south west, inner sep=0] (image) at (9.9,6.7)
{\includegraphics[trim={0 0 0 0},clip,width=0.22\textwidth]{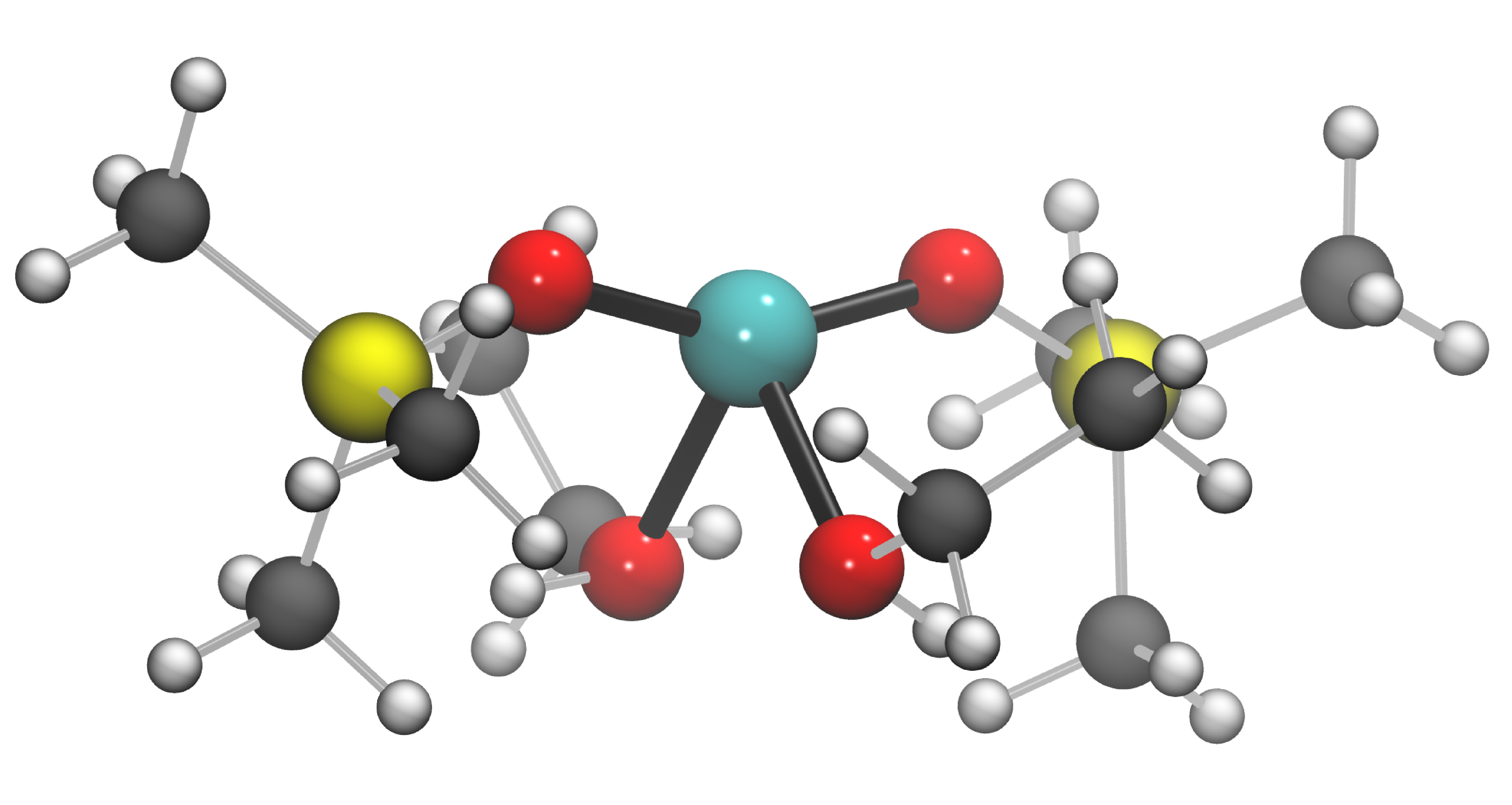}};
\node at (10,9.6) {\large{(6)}};

\node at (-4,18) {\large{$\mathbf{A}$}};
\node at (-4,13.7) {\large{$\mathbf{B}$}};
\node at (-4,9.5) {\large{$\mathbf{C}$}};
\node at (-4,5.5) {\large{$\mathbf{D}$}};
\node at (9,18) {\large{$\mathbf{E}$}};
\node at (9,5.5) {\large{$\mathbf{F}$}};
\def\s{7}
\def\sr{11}


\draw [ultra thick, ->,>=stealth] (-1.5+\s,2) -- (-1.5+\s,5.5);
\node [rotate=90] at (-1.8+\s,3.7) {Energy};


\draw [ultra thick] (-1+\s,2) -- (0+\s,2); 
\draw [ultra thick] (-1+\s,3.5) -- (0+\s,3.5); 
\draw [ultra thick] (1+\s,2) -- (2+\s,2);
\draw [ultra thick] (1+\s,3.5) -- (2+\s,3.5); 
\draw [ultra thick] (0+\s,4.5) -- (1+\s,4.5); 

\draw [dotted] (0+\s,3.5) -- (1+\s,3.5); 
\draw [dotted] (0+\s,2) -- (1+\s,2); 

\node at (-0.5+\s,1.5) {d$_{x^2-y^2}$};
\node at (-0.5+\s,4) {d$_{xz}$};
\node at (1.5+\s,1.5) {d$_{xy}$};
\node at (1.5+\s,4) {d$_{yz}$};
\node at (0.5+\s,5) {d$_{z^2}$};


\draw [ultra thick] (-1+\sr,2) -- (0+\sr,2); 
\draw [ultra thick] (-1+\sr,4.5) -- (0+\sr,4.5); 
\draw [ultra thick] (1+\sr,2) -- (2+\sr,2);
\draw [ultra thick] (1+\sr,4.5) -- (2+\sr,4.5); 
\draw [ultra thick] (0+\sr,3.5) -- (1+\sr,3.5); 

\draw [dotted] (0+\sr,2) -- (1+\sr,2); 
\draw [ <->,>=stealth] (0.5+\sr,2) -- (0.5+\sr,3.5);
\node at (0.2+\sr,2.7) {$\Delta$};

\node at (-0.5+\sr,1.5) {d$_{x^2-y^2}$};
\node at (-0.5+\sr,5) {d$_{xz}$};
\node at (1.5+\sr,1.5) {d$_{z^2}$};
\node at (1.5+\sr,5) {d$_{yz}$};
\node at (0.5+\sr,4) {d$_{xy}$};

\end{tikzpicture}
\caption{\textbf{A}: distribution of the computed anisotropy $D$ values for \textbf{Set-1} (green) and \textbf{Set-2} (orange). The density of states for \textbf{Set-2} is multiplied by 50 to facilitate the visualization. Black bars correspond to literature values of  [Co(II)(C(SiMe$_2$ONaph)$_3$)$_2$] (Me=methyl, Naph=naphthyl) \cite{Long2018},  [Co(II)(sIPr)NDmp] (Dmp=2,6-dimesitylphenyl) \cite{Yao2016}, (Ph$_4$P)$_2$[Co(II)(C$_3$S$_5$)$_2$]  \cite{Fataftah2014},[Co(II)(IPr)NDmp] (Dmp=2,6-dimesitylphenyl) \cite{Yao2016}, [Co(II)(cyIPr)NDmp] (Dmp=2,6-dimesitylphenyl) \cite{Yao2016}, (HNEt$_3$)$_2$[Co(II)(L)$_2$] (H$_2$L=1,2-bis(methanesulfonamido)benzene) \cite{Rechkemmer2016}, from the most negative to the least negative, respectively. In the inset, only values of D $<$ -100 cm$^{-1}$ are reported and no multiplication factor has been applied for \textbf{Set-2}. 
\textbf{B}: computed values of rhombic parameter $E$ with respect to anisotropy $D$. Green dots is used for \textbf{Set-1} and orange dots for \textbf{Set-2}.   
\textbf{C}: expectation value of the $z$-component of the orbital angular momentum $L$ as a function of the computed anisotropy $D$. Values are reported for \textbf{Set-1} (green) and \textbf{Set-2} (orange). \textbf{D}: the relative percentage of each reference polyhedron in \textbf{Set-1} across different $D$ ranges is represented as follows: yellow for square-planar, orange for seesaw, sea green for octahedral, and light cyan for tetrahedral. 
\textbf{E}: representative geometries for different reference coordination in the two sets. (1) linear, (2) square planar, (3) seesaw, (4) octahedral with large negative $D$, (5) octahedral with large positive $D$, (6) tetrahedral,. Color code: cyan for Co, blue for N, red for O, grey for C, white for H, orange for Se, yellow for Si, green for F, magenta for S, and purple for P. \textbf{F}: molecular orbital diagram for linear coordination (left) and tetrahedral coordination (right).}
\label{D}
\end{figure*}

\textbf{Magnetic anisotropy.} Co(II) ions possess seven electrons in their $d$ shell, leading to a high spin ground state with $S=3/2$. The $2S+1=4$ states of such a spin, split in energy by the presence of a crystal field and spin-orbit coupling, can be described by the conventional spin Hamiltonian
\begin{equation}   \hat{H}_S=D\hat{S}^2_z+E(\hat{S}^2_x-\hat{S}^2_y)\:,
\label{SH}
\end{equation}
where $D$ and $E$ are the axial and rhombic zero-field splitting parameters, respectively. Magnetic anisotropy is defined as the difference in energy between the first and second Kramers doublets. For vanishing values of $E$, these states correspond to the maximal and minimal projections of the spin along the $z$ axis, $M_S=\pm 3/2$ and $M_S=\pm 1/2$, respectively, and their energy difference is $2D$. In the presence of negative values of the parameter $D$, the energy of the maximally large projections of the spin along the $z$ axis ($M_S=\pm 3/2$) is stabilized, leading to an easy-axis magnetic anisotropy, known to be favourable for single-molecule magnets\cite{Raza2023,Gatteschi2006-hw}. Positive values of $D$ correspond to easy-plane magnetic anisotropy. The parameter $E$ leads to the mixing of different values of $M_S$, promoting quantum tunnelling of the magnetization and undercutting the benefits of large $D$ values\cite{Sirenko2024}. The requirement of maximising $|D/E|$ is therefore key to designing single-molecule magnets.

The computed values of $D$ are reported in Fig. \ref{D}A. For \textbf{Set-1}, the distribution of values of $D$ shows two pronounced peaks around $D\sim$ 0 cm$^{-1}$, indicating that a random selection of ligands is likely to result in complexes with small non-zero magnetic anisotropy. In contrast, for \textbf{Set-2}, the majority of compounds exhibit large negative $D$ values and no positive values. The large size of \textbf{Set-1} allows to sample the tails of the distribution which are found to extend to significantly large values of $|D|$. The inset of Fig. \ref{D}A reveals 196 complexes from \textbf{Set-1} and 47 from \textbf{Set-2} with magnetic anisotropy $D<-100$ cm$^{-1}$, which is comparable to the best compounds synthesized to date\cite{Long2018,Yao2016,Fataftah2014,Rechkemmer2016}. Further exploration in the region of low $D$ values reveals 22 compounds from \textbf{Set-1} and 27 compounds from set \textbf{Set-2} with magnetic anisotropy smaller than -200 cm$^{-1}$, comparable to the record values of this property for Cobalt\cite{Long2018,Yao2016}. Interestingly, a significant number of compounds with large positive values of $D$ have been detected in \textbf{Set-1}. Fig. \ref{D}A also reports the value of $D$ for the six complexes with the highest negative magnetic anisotropy reported in the literature. Four of these compounds exhibit linear coordination of the central Co(II) with pseudo-$D$ values of -221\cite{Long2018}, -206\cite{Yao2016}, -154\cite{Yao2016}, and -140 cm$^{-1}$ \cite{Yao2016}. The other two compounds instead features distorted tetrahedral coordination and $D \sim$ -113\cite{Rechkemmer2016} and -161 cm$^{-1}$\cite{Fataftah2014}. The values of the rhombic parameter $E$ are presented in Fig. \ref{D}B, where the theoretical limit $|E/D| < 1/3$ is clearly visible. Interestingly, the parameter $E$ spans a large range of values in both \textbf{Set-1} and \textbf{Set-2} except for $D<-200$ cm$^{-1}$, suggesting that maximising $D/E$ is possible.

\textbf{Angular momentum, coordination geometries and molecular orbitals.}  To test the hypothesis that large magnetic anisotropy emerges from unquenched angular momentum\cite{Long2018,Atanasov2013,Zadrozny2013,Zadrozny2013_2}, we calculated the expectation value of the operator $L_z $ for each compound. These results are presented in Fig. \ref{D}B. A consistent correlation between magnetic anisotropy and $\langle L_z \rangle$ is observed across all compounds, reaching the maximum theoretical value of 3 for the most anisotropic ones. In light of the observed relationship between $\langle L_z \rangle$ and magnetic anisotropy, we now examine the geometric arrangement of atoms surrounding the magnetic ion, as well as the resulting energy distribution and occupation of the 3$d$ molecular orbitals (MOs). In the following, we first focus on \textbf{Set-2}, followed by the analysis of the compounds in \textbf{Set-1}.

The vast majority of compounds prepared in linear coordination maintain the angle $\theta$ between the Co ion and the two coordinated ligands close to 180$^\circ$ (see Fig. \ref{D}E-(1)). It is possible to understand why this geometry supports large values of $\langle\hat{L}_z\rangle$ by considering the MO diagram for a linear compound, as shown in the left panel of Fig. \ref{D}F. The 3$d$ electronic states are ordered as $E(d_{z^2}) > E(d_{xz}, d_{yz}) > E(d_{x^2-y^2}, d_{xy})$. The degeneracy of the $d_{x^2-y^2}$ and $d_{xy}$ orbitals ($m_l = \pm 2$), and the $d_{xz}$ and $d_{yz}$ orbitals ($m_l = \pm 1$), leads to unquenched orbital angular momentum if there is an odd number of electrons in one of these two manifolds. Following the Aufbau principle to populate the 3$d$ orbitals, the resulting occupation would be $(d_{z^2})^1(d_{xz}, d_{yz})^2(d_{x^2-y^2}, d_{xy})^4$, leading to a final $\langle L_z \rangle = 0$. However, a non-Aufbau filling of 3$d$ orbitals would result in the configuration $(d_{z^2})^1(d_{xz}, d_{yz})^3(d_{x^2-y^2}, d_{xy})^3$, yielding $\langle L_z \rangle = 3$. In the set of compounds studied here, large values of $\langle L_z \rangle$ are indeed found for nearly linear geometries, leading to the conclusion that the large magnetic anisotropy for these compounds is due to the non-Aufbau occupation of the 3$d$ orbitals. A closer inspection of the trend of $\theta$ (reported in Fig. S1), reveals that deviations from linearity correlate with a decrease in the computed $D$ and $\langle L_z \rangle$ values. This decrease is explained by a loss of degeneracy between the pairs of orbitals $d_{x^2-y^2}$-$d_{xy}$ and $d_{xz}$-$d_{yz}$. \\

Interestingly, large values of $\langle L_z \rangle$ and $D$ surpassing those reported in the literature have been detected in compounds with coordination higher than two in \textbf{Set-1}. Despite the initial complexes being prepared in tetrahedral coordination, DFT optimization leads to equilibrium geometries that significantly deviate from it. In particular, we note that coordination larger than four is also established because some ligands possess a second donor atom despite being singly coordinated in the originally deposited crystallographic structure. In some cases, DFT geometrical optimization then leads to compounds with coordination six, where two ligands act as bi-dentate.

The first coordination shell geometry of each compound in \textbf{Set-1} is interpreted in terms of ideal polyhedrons as detailed in the Methods section\cite{Casanova2004,Cirera2006}. In the final optimized geometries of \textbf{Set-1}, compounds close to square planar, tetrahedral, seesaw, and octahedral are present and Fig. \ref{D}D reports the relative percentage of each reference polyhedron in different windows of $D$ values. In particular, we observe that in the range of $D$ values smaller than -150 cm$^{-1}$, the majority of the complexes relax to geometries close to square planar, seesaw or octahedral coordination (see Figs. \ref{D}E-(2-4)). Further inspection of these compounds reveals the establishment of two sets of metal-to-ligand distances in the same compounds, with two ligands lying along the same direction in a pseudo-linear fashion and exhibiting short bond lengths and the remaining ligands lying in a perpendicular plane and showing longer bond lengths. These geometries lead to strong axial anisotropy and the emergence of pseudo-linear MO configuration, as schematically reported in Fig. S2. In contrast with the case of linear two-coordinated compounds, the degeneracy of the $d_{xz}$, $d_{yz}$ and $d_{x^2-y^2}$, $d_{xy}$ orbitals is partially lifted by the presence of the equatorial ligands. Nevertheless, the weak interaction between the metal and the equatorial ligands results in a small energy splitting between these orbitals such that unquenched orbital angular momentum is still supported. 
Interestingly, when the four equatorial donor atoms of octahedral geometries lie close to the metal and the metal-axial bond length is elongated (see Fig. \ref{D}E-(5)), significant in-plane anisotropy is observed.
Finally, tetrahedral coordination remains favourable for compounds with magnetic anisotropy values ranging from -150 and 50 cm$^{-1}$. In this case, large magnetic anisotropy is associated with a small energy gap between the $d_{x^2-y^2}$ and $d_{xy}$ orbitals \cite{Rechkemmer2016}. As illustrated in the right panel of Fig. \ref{D}F, as the separation $\Delta$ between $E(d_{xy})$ and $E(d_{x^2-y^2})$ vanishes, $\langle L_z \rangle$ approaches the value of two and magnetic anisotropy becomes large. Computed values of $\Delta$ are reported in Fig. S3.

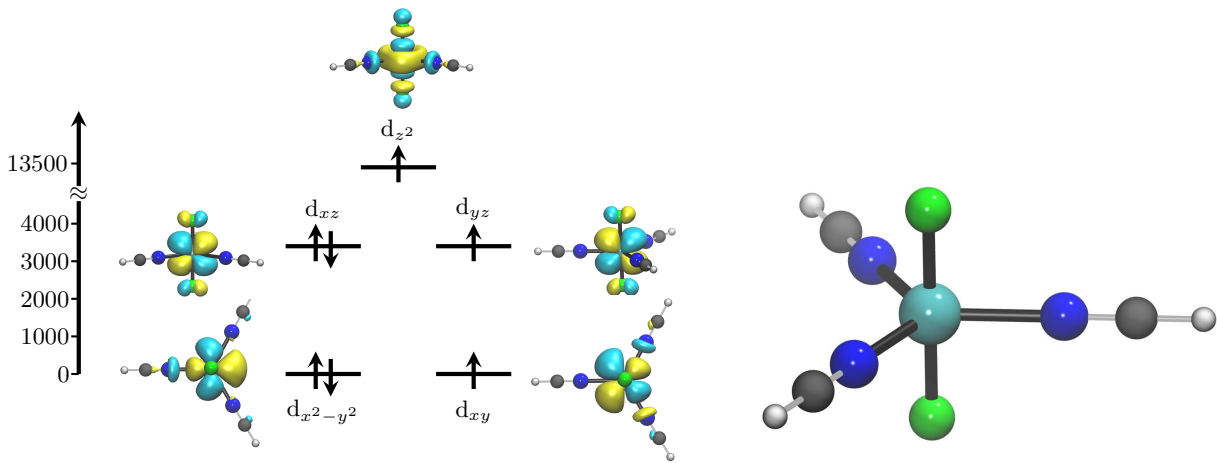
\begin{figure*}

\begin{tikzpicture}

\def\s{-0.25}
\def\sr{7.5}

\node[anchor=south west, inner sep=0] (image) at (5,1.0+\sr)
{\includegraphics[clip,width=0.35\textwidth]{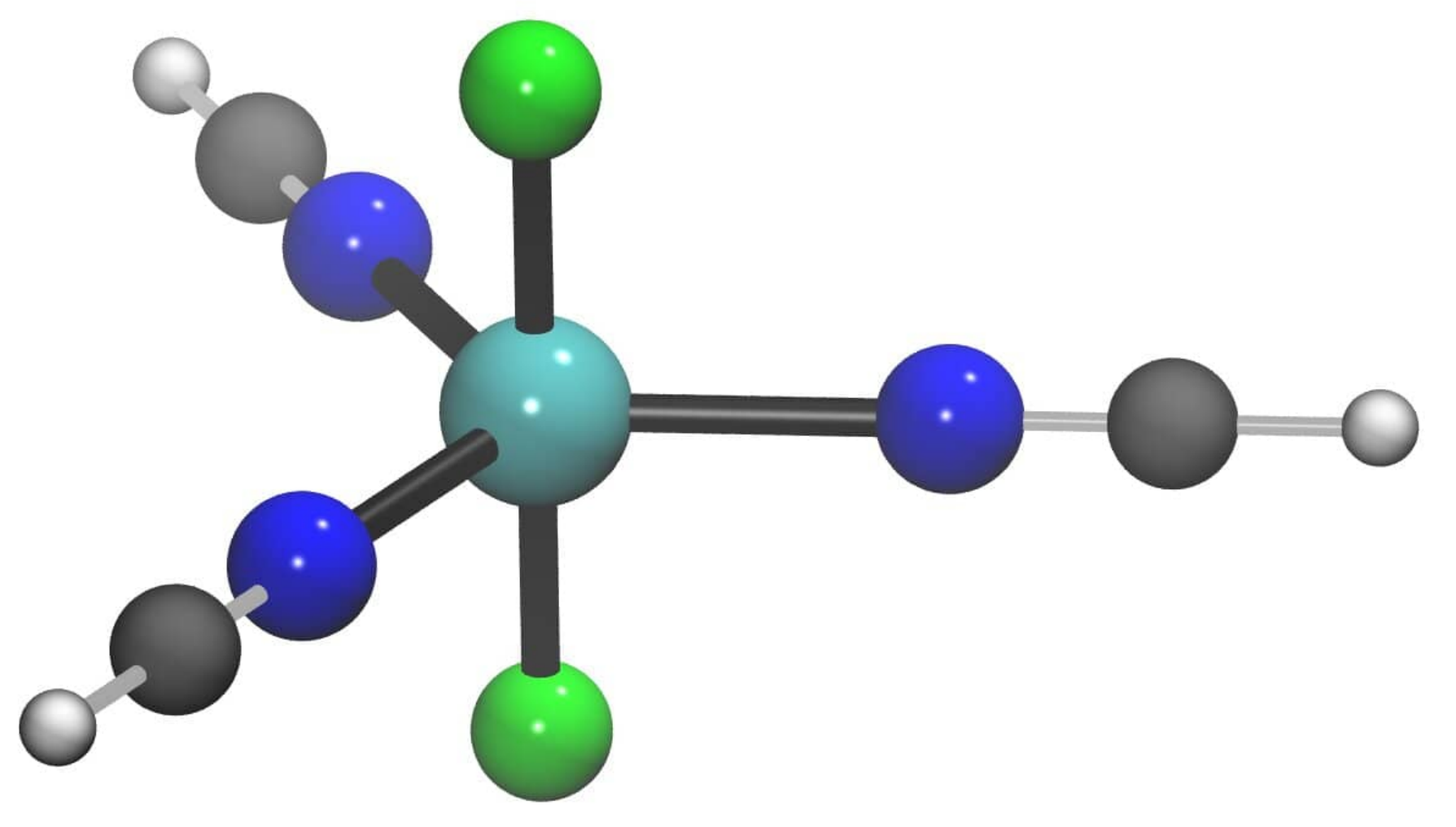}};
\node[anchor=south west, inner sep=0] (image) at (-3.5,1.0+\sr)
{\includegraphics[clip,width=0.11\textwidth]{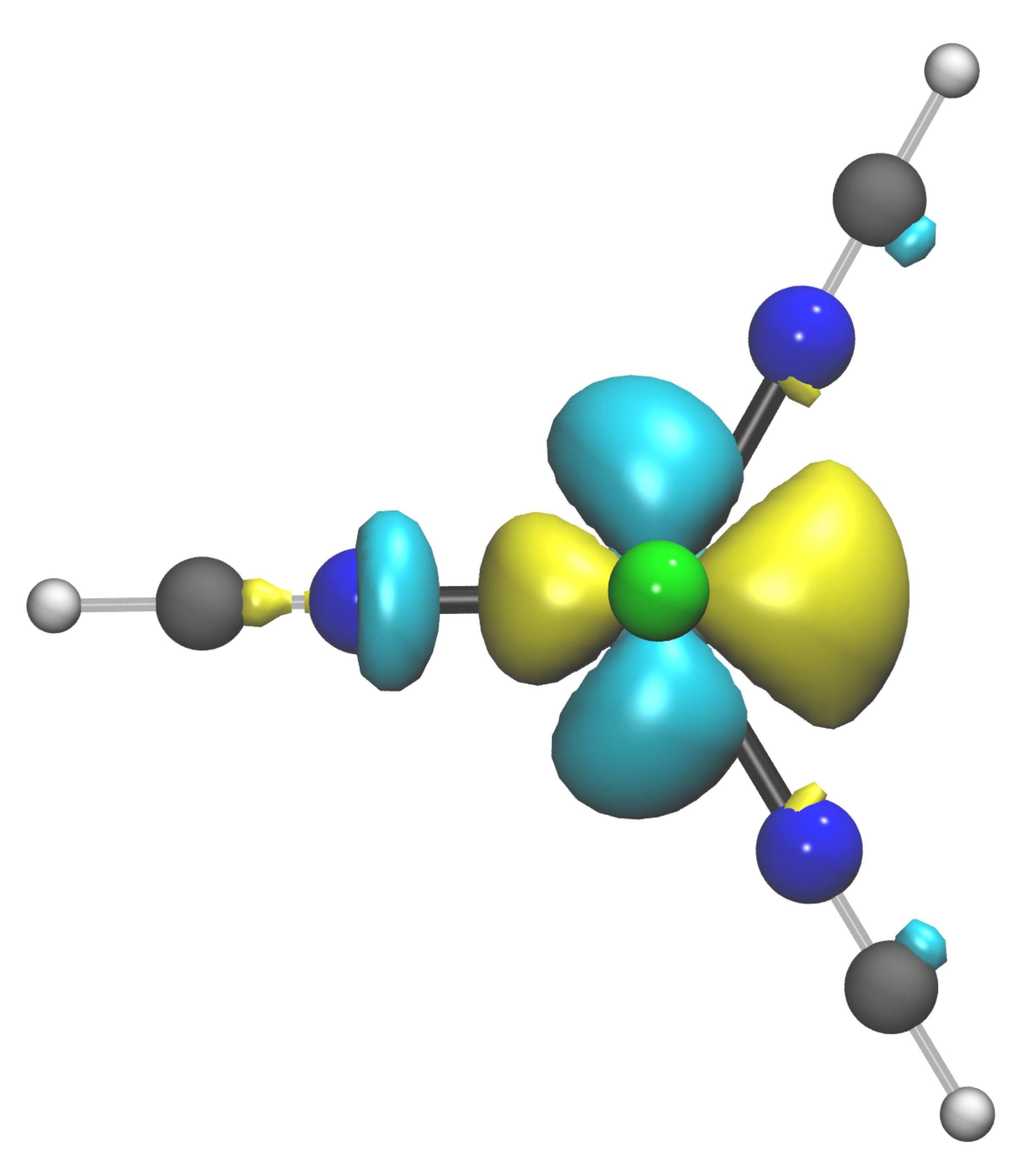}};
\node[anchor=south west, inner sep=0] (image) at (-3.5,3.0+\sr)
{\includegraphics[clip,width=0.11\textwidth]{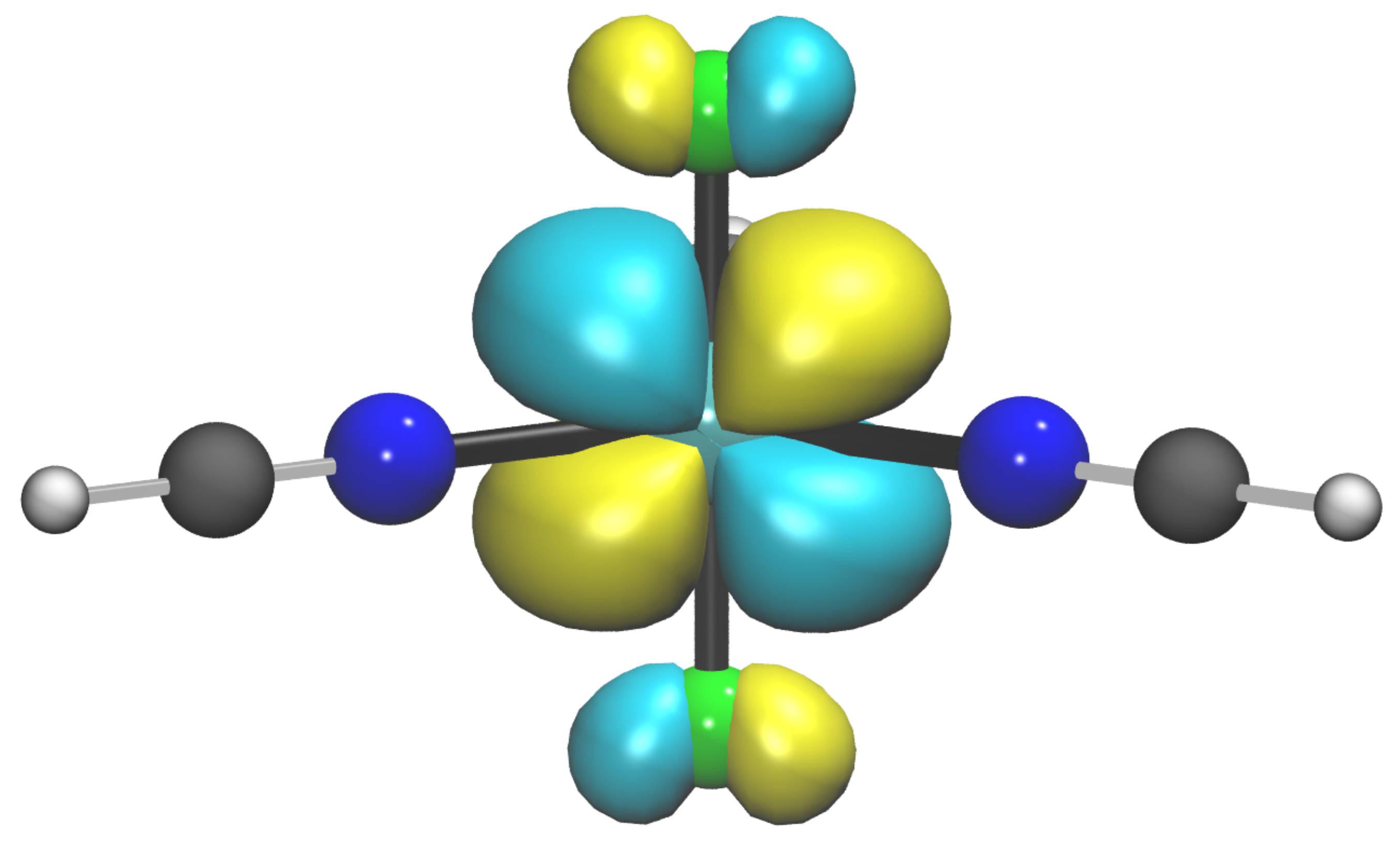}};
\node[anchor=south west, inner sep=0] (image) at (2,3.0+\sr)
{\includegraphics[clip,width=0.11\textwidth]{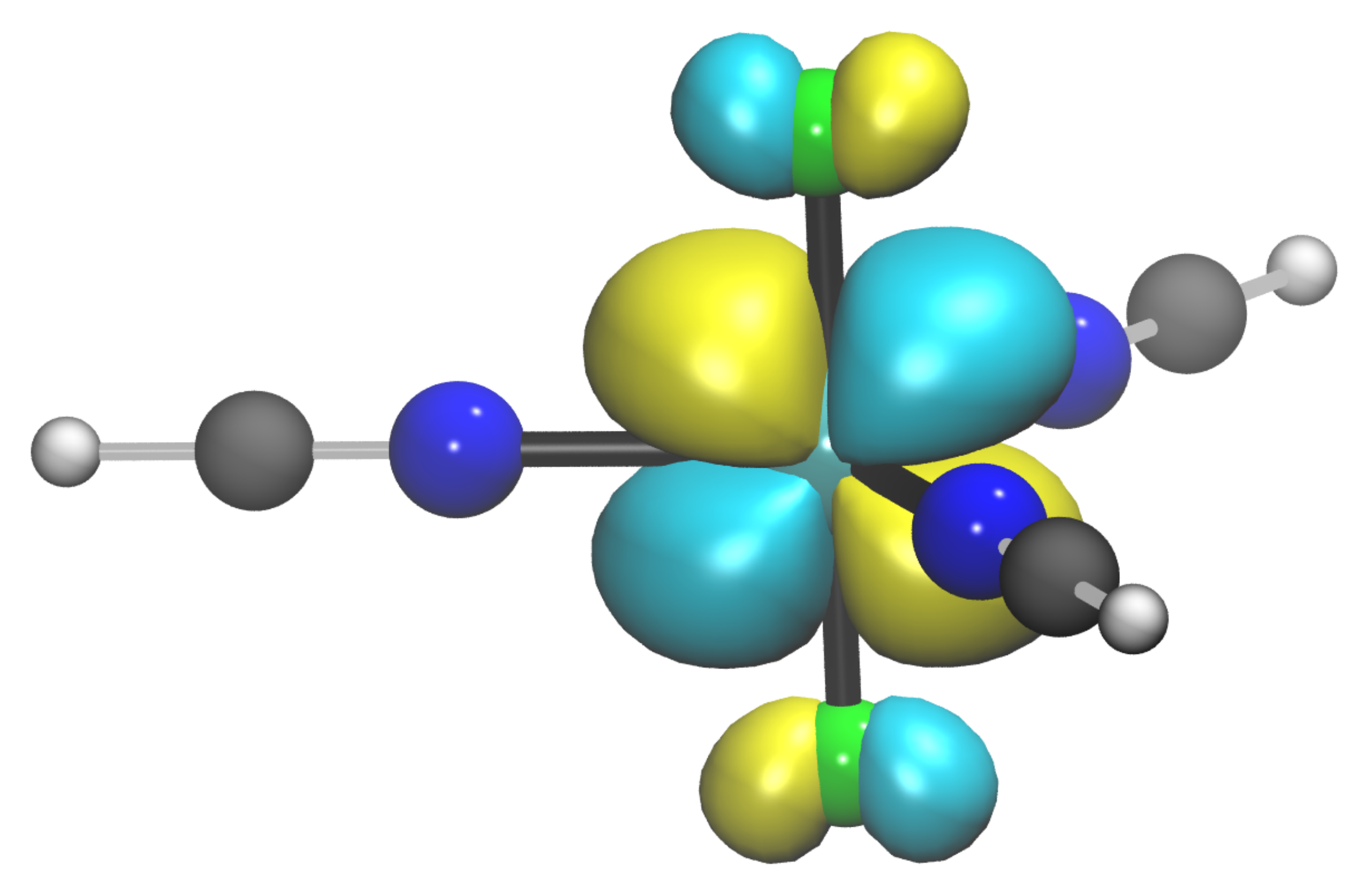}};
\node[anchor=south west, inner sep=0] (image) at (2,0.8+\sr)
{\includegraphics[clip,width=0.11\textwidth]{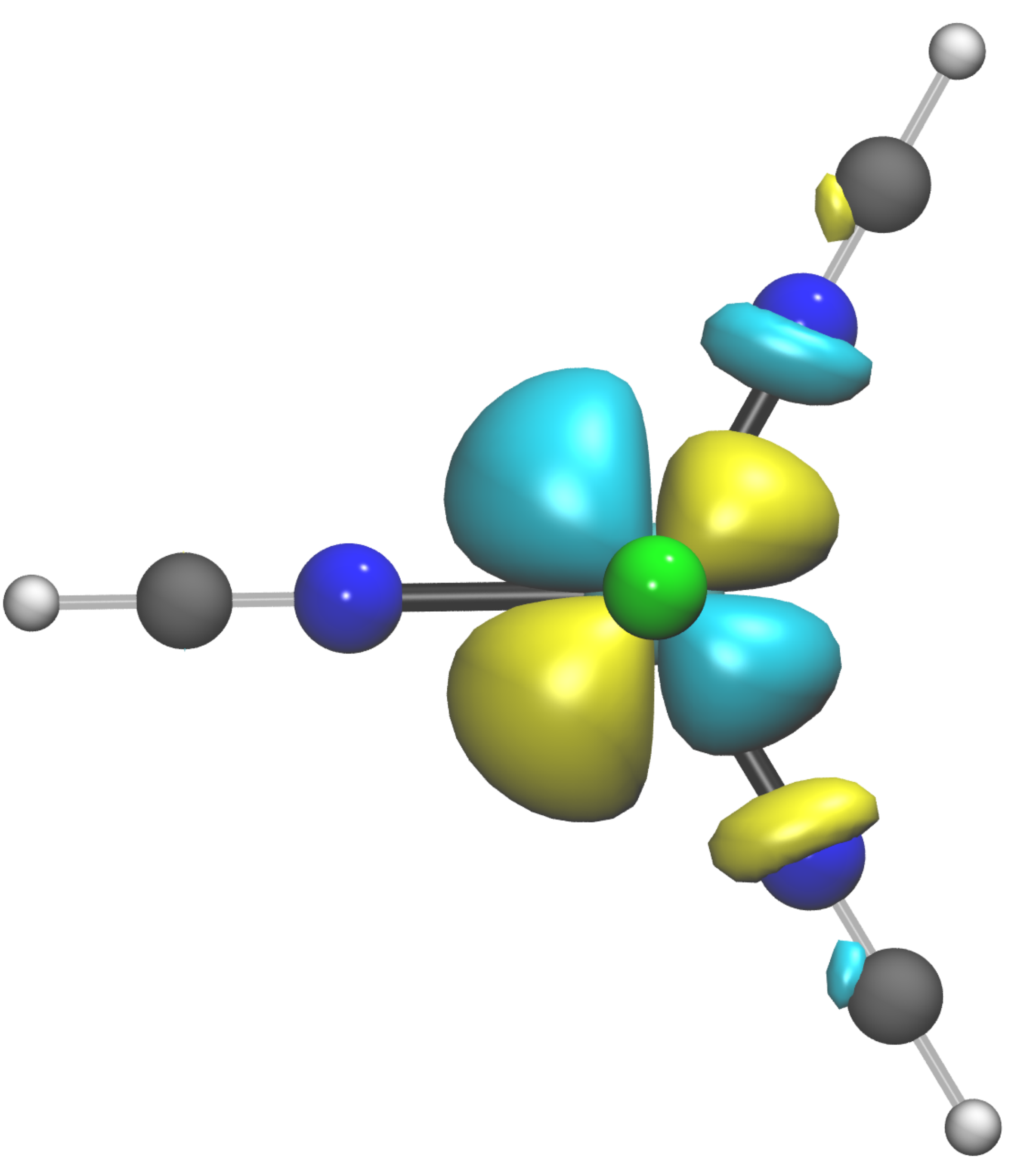}};
\node[anchor=south west, inner sep=0] (image) at (-0.7,5.5+\sr)
{\includegraphics[clip,width=0.11\textwidth]{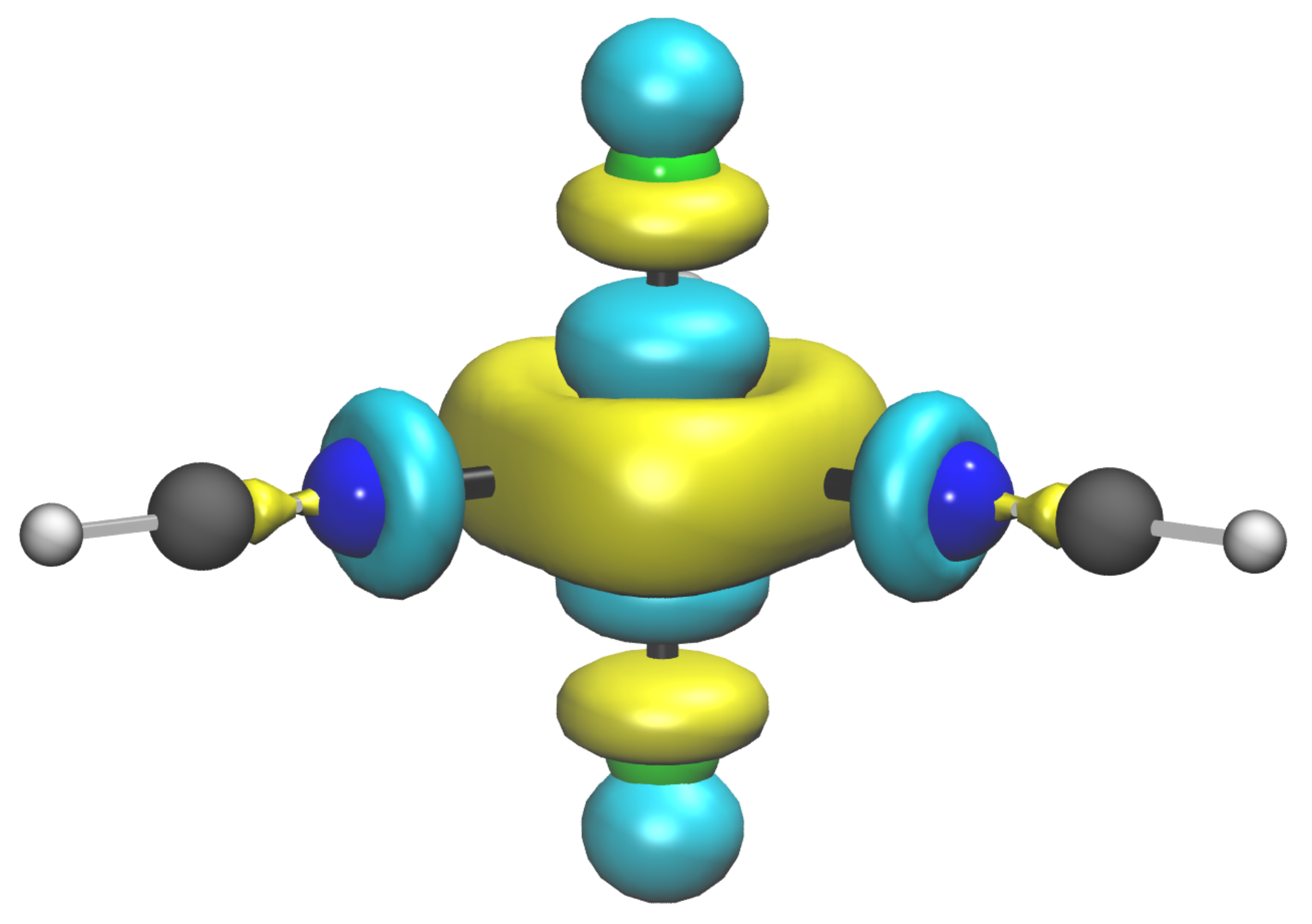}};


\draw [ultra thick] (-3.75+\s,2+\sr) -- (-3.75+\s,4.3+\sr);
\node at (-3.75+\s,4.43+\sr) {$\sim$};
\node at (-3.75+\s,4.35+\sr) {$\sim$};
\draw [ultra thick, ->,>=stealth] (-3.75+\s,4.5+\sr) -- (-3.75+\s,5.5+\sr);

\draw [thick] (-3.75+\s,2+\sr) -- (-3.85+\s,2+\sr);
\node at (-3.95+\s,2+\sr) {0};
\draw [thick] (-3.75+\s,2.5+\sr) -- (-3.85+\s,2.5+\sr);
\node at (-4.2+\s,2.5+\sr) {1000};
\draw [thick] (-3.75+\s,3+\sr) -- (-3.85+\s,3+\sr);
\node at (-4.2+\s,3+\sr) {2000};
\draw [thick] (-3.75+\s,3.5+\sr) -- (-3.85+\s,3.5+\sr);
\node at (-4.2+\s,3.5+\sr) {3000};
\draw [thick] (-3.75+\s,4+\sr) -- (-3.85+\s,4+\sr);
\node at (-4.2+\s,4+\sr) {4000};
\draw [thick] (-3.75+\s,4.8+\sr) -- (-3.85+\s,4.8+\sr);
\node at (-4.3+\s,4.8+\sr) {13500};


\draw [ultra thick] (-1+\s,2+\sr) -- (0+\s,2+\sr); 
\draw [ultra thick, ->,>=stealth] (-0.6+\s,1.8+\sr) -- (-0.6+\s,2.3+\sr);
\draw [ultra thick, <-,>=stealth] (-0.4+\s,1.7+\sr) -- (-0.4+\s,2.2+\sr);
\draw [ultra thick] (-1+\s,3.7+\sr) -- (0+\s,3.7+\sr); 
\draw [ultra thick, ->,>=stealth] (-0.6+\s,3.5+\sr) -- (-0.6+\s,4.0+\sr);
\draw [ultra thick, <-,>=stealth] (-0.4+\s,3.4+\sr) -- (-0.4+\s,3.9+\sr);
\draw [ultra thick] (1+\s,2+\sr) -- (2+\s,2+\sr);
\draw [ultra thick, ->,>=stealth] (1.5+\s,1.8+\sr) -- (1.5+\s,2.3+\sr);
\draw [ultra thick] (1+\s,3.7+\sr) -- (2+\s,3.7+\sr); 
\draw [ultra thick, ->,>=stealth] (1.5+\s,3.5+\sr) -- (1.5+\s,4.0+\sr);
\draw [ultra thick] (0+\s,4.75+\sr) -- (1+\s,4.75+\sr); 
\draw [ultra thick, ->,>=stealth] (0.5+\s,4.55+\sr) -- (0.5+\s,5.05+\sr);

\node at (-0.5+\s,1.5+\sr) {d$_{x^2-y^2}$};
\node at (-0.5+\s,4.2+\sr) {d$_{xz}$};
\node at (1.5+\s,1.5+\sr) {d$_{xy}$};
\node at (1.5+\s,4.2+\sr) {d$_{yz}$};
\node at (0.5+\s,5.25+\sr) {d$_{z^2}$};

\end{tikzpicture}
\caption{MOs diagram, orbitals occupation, and optimised DFT geometry for the proposed trigonal-bipyramidal compound. CASSCF orbitals are shown next to each 3d electronic level. Color code: cyan for Co, blue for N, grey for C, white for H, and green for F. }
\label{MOs}
\end{figure*}

\textbf{Elemental analysis.} Most of the ligands in the initial set bind to Co with nitrogen (32.7 \%), carbon (27.9 \%), and oxygen (21.6 \%). This distribution is reflected in the composition of the first-coordination sphere of complexes from \textbf{Set-1}, where 8,694 contain nitrogen, 7,384 contain carbon, and 6,415 contain oxygen.
Interestingly, when the analysis is repeated considering only compounds with computed $D$ values smaller than -100 cm$^{-1}$, the statistic is reversed, with most of the 197 compounds containing oxygen (146), followed by nitrogen (98) and carbon (24). To understand the role of oxygen in maximizing magnetic anisotropy, we need to consider the MO diagram in Fig. \ref{D}F.
Such a configuration can be obtained in the case of strong $\pi$-donor ligands in the axial direction and weak bonding ligands in the equatorial plane. This situation ensures an energy destabilization of the non-bonding $d_{xz}$, $d_{yz}$ orbitals with respect to the $d_{x^2-y^2}$, $d_{xy}$ orbitals. Highly electronegative elements, such as oxygen, are optimal donor atoms to enhance the metal-ligand $\pi$ interaction because of the presence of multiple lone electron pairs that can establish chemical bonds with the magnetic ion. Ligands of this type have been found in around 65 \% of the compounds with magnetic anisotropy smaller than -200 cm$^{-1}$. Oxygen, Carbon and Nitrogen are all present in the first coordination shell of compounds in \textbf{Set-2}, but the small size of the latter does not allow the extraction of statistically significant information about their relative importance.

\section*{Discussion}

The community of molecular magnetism has been actively pursuing the synthesis of Co single-molecule magnets for over two decades and many different chemical strategies have been successfully pursued\cite{KumarSahu2023}. Very often these efforts are driven by assuming the existence of a direct mapping between coordination geometry and magnetic properties. Our results show that whilst coordination geometry indeed plays a key role in shaping the magnetic anisotropy of an ion, it cannot be separated by other key factors such as the cross-correlations among the ligands' different ability to form coordination bonds. Thanks to an unprecedented large-scale screening of Co compounds, we have here uncovered several examples where prominent values of zero-field splitting are achieved in unexpected coordination geometries, ranging from square planar and see-saw to octahedral.

We anticipate that the synthesis of the compounds reported here will come with significant challenges, as is common for computational design studies of this kind\cite{Szymanski2023,PyzerKnapp2022,Oganov2019}. In particular, it is currently beyond the state of the art to systematically explore the chemical stability and synthetic feasibility of coordination compounds against the immense number of chemical environments, possible crystal structures and competitive reactions that can take place in real experiments. Nonetheless, a very clear and actionable design rule emerges from our study: a maximal value of angular momentum, and therefore of axial magnetic anisotropy, can be achieved with pseudo-linear geometries, where two strong donors are aligned along the same direction and only weak donors are present in the equatorial plane. To further support the universality of our claims, we assemble a complex with trigonal pyramidal geometry where axial and equatorial ligands are assigned based on our design rule. Fig. \ref{MOs} reports the optimized structure and MOs for a complex with two fluorine ions placed on the main symmetry axis and three NCH ligands on the equatorial positions. As expected from the position of these ligands on the spectrochemical series, the geometry shows a distance of 1.84 \AA$ $ between the apical F$^-$ and the Co(II) ion, while the distance between the NCH ligands and the metal is 2.14 \AA$ $. Complete intra-manifold degeneracy is achieved for $(d_{xz}, d_{yz})$ and $(d_{x^2-y^2}, d_{xy})$, respectively, thanks to the $C_3$ symmetry axis. The result is a computed magnetic anisotropy of 253 cm$^{-1}$, the largest theoretical value reported so far. The anisotropy of Co in such coordination geometry had been studied experimentally before, but only marginal values of $D$ had been achieved due to the lack of the specific tuning of the ligands' reciprocal donor strength that emerged from this study\cite{Zahradnkov2023,Brachakov2020}. Although linear two-coordinate compounds statistically exhibit a larger magnetic anisotropy than any other coordination, compounds with higher coordination numbers have the potential to lead to molecules with improved chemical stability, a \textit{sine qua non} condition for any application of these compounds. A similar strategy has recently emerged in the field of air-stable lanthanide single-molecule magnets\cite{Canaj2019,Li2019,Zhu2021}, supporting the present analysis.  

In conclusion, we have developed a high-throughput multi-reference ab initio framework to automatically assemble over 15,000 coordination compounds of Co(II) and compute their magnetic anisotropy. We discovered tens of molecules with record values of magnetic anisotropy and a general new chemical design rule that does not invoke low coordination numbers. We expect that these results will facilitate further large-scale explorations of magnetic molecules' chemical space as well as the experimental characterization of a new class of Cobalt compounds with unprecedented properties.

\section*{Methods}

\noindent
\textbf{Ligands database generation.} 
A total of 47,427 structures containing Co were downloaded from CSD on April 2023. Cif files presenting structural disorder are discarded. The remaining cif files are then converted to xyz with the software Atomsk \cite{Atomsk}. The xyz files are further processed with the tool cif2xyz, available as part of the software MolForge, available at https://github.com/LunghiGroup/MolForge/. The tool FindMols, also available in MolForge, is then used to individuate molecular entities inside the unit cells and to remap them across the periodic boundary conditions. Crystals containing a unique single-ion molecule of Co are retained, leaving 35,632 entries. For each molecule, the ligands are identified with the tool FindMols. The brute formulae of ligands are compared to remove duplicates. We further differentiate ligands by the number of donor atoms and select the monodentate ones. This leaves us with a list of 1,424 unique ligands. Bispectrum components are used to represent the structure of the donor atom's chemical environmet in each ligand. Bispectrum components\cite{Bartk2013} are built with a cut off distance of 4.1 \AA$ $ and order $2J=8$. PCA is then performed on bispectrum components to obtain a 2D representation of the structural similarity of ligands' donor atoms. Then, FPS was used to select 208 points on this map that best represented the space, i.e. 208 ligands that are most chemically diverse out of the initial list of ligands. The charge of each ligand is determined using DFT with the same functional and basis set employed for the geometrical optimization (see below). Without performing geometrical optimization, single-point calculations are conducted by varying the total charge of the compound from 2- to 2+ and adjusting the spin multiplicity accordingly. The charge is determined based on the state with the lowest total energy. No open-shell solutions are found to be the most stable configuration for the analyzed ligands. In the case of carbon donor atoms, carbanion solutions are selected by default. 

\vspace{0.2cm}
\noindent
\textbf{Generation of Co(II) compounds.} 
Starting from the geometries of the selected 208 ligands, along with their associated charges and docking atoms, the software molSimplify \cite{molSimplify} is used to generate the initial geometries of the Co(A)$_2$(B)$_2$ (\textbf{Set-1}) and Co(A)$_2$ (\textbf{Set-2}) compounds by exploring all possible combinations of the ligands within each set. The compounds with a brute formula Co(A)$_4$ are not considered as they are not expected to lead to significant symmetry breaking to support large magnetic anisotropy. For \textbf{Set-1}, we use the template for tetrahedral coordination, while for \textbf{Set-2}, the linear coordination template is applied. Aside from the steric repulsion minimization performed by the molSimplify routine, no additional force-field options are used.

Cobalt is considered to be in its high spin configuration ($S=3/2$) in all subsequent calculations. Preoptimization of the generated geometries is performed only for compounds in \textbf{Set-1} using semi-empirical tight-binding DFT (TB-DFT) with the GFN2-xTB method\cite{Bannwarth2020,GFN2}, as implemented in the ORCA 5\cite{neese2020orca} software. The optimization threshold is set to $10^{-6}$ Ha (tight level). Structures for which TB-DFT optimization fails to converge after 100 cycles are replaced with the initial structures generated using molSimplify. Finally, all the structures from both sets are optimized at the DFT level using ORCA 5\cite{neese2020orca}. We employ the BP86 functional\cite{Becke1988,Perdew1986} with the addition of dispersion correction at the D3-BJ level\cite{Grimme2010}. All parameters for dispersion corrections are kept at their default values. The def2-TZVPP basis set is used for all atoms. Calculations that do not converge after 500 geometry optimization cycles are discarded. At the end of the optimization run, unrestricted natural orbitals (UNO) are generated for the subsequent multi-reference calculations.

\vspace{0.2cm}
\noindent
\textbf{Multi-reference calculations.} Multireference calculations are performed using the State-Average Complete Active Space Self Consistent Field (CASSCF) theory as implemented in the software ORCA 5\cite{neese2020orca}. The active space used to build the CASSCF wavefunction is (7,5), i.e., seven electrons in five 3$d$-orbitals. The right active space is automatically selected using the Löewdin orbital composition to assess the percentage of Co $d$ orbitals in each molecular orbital (MO). If any of the last five occupied orbitals have less than 30 $\%$ $d$ character, the highest-energy occupied orbitals (outside these last five) with more than 30 $\%$ $d$ character are identified and substituted into the active space. The CASSCF calculation is then run with these swapped orbitals. This procedure is initially applied to the UNO orbitals from DFT and then to each subsequent CASSCF run until the active space has the correct composition. The state average procedure is performed using 10 quartet states ($2S+1=4$) and 40 doublet states ($2S+1=2$). Mean-field spin-orbit coupling operator, along with Quasi Degenerate Perturbation Theory (QDPT), is employed to account for the mixing of spin-free states. The spin Hamiltonian reported in Eq. \ref{SH} is built using the lowest two Kramers doublets. The Douglas-Kroll-Hess (DKH) scalar relativistic correction is applied to the electronic Hamiltonian, with picture change effects considered up to second order to include DKH corrections in the spin-orbit coupling operator. We use the DKH-def2-TZVPP basis set for all atoms, except for Sn, for which the SARC-DKH-TZVPP basis set is employed. The $\Delta$ values reported in Fig. \ref{D} F were extracted from the eigenvalues of the ligand field one-electron matrix built using Ab Initio Ligand Field Theory (AILFT) on top of the CASSCF orbitals\cite{atanasov2011,Lang2020}. Total energy differences between low-spin ($S=1/2$) and high-spin ($S=3/2$) configurations are reported in Fig. S4.

\vspace{0.2cm}
\noindent
\textbf{Assignment of reference polyhedron.} To connect a given geometry with a reference polyhedron, we start by analyzing the number and composition of the first coordination sphere. This involves examining the distribution of distances between the Co ion and all other elements in the compounds, as shown in Fig. S5. For each element $i$, we identify the position of the first and the second peak in this distribution as $d_{1,i}$ and $d_{2,i}$, respectively. These values are reported in Table S1. To determine if a specific atom of species $i$ is part of the first coordination sphere in a given geometry, we check if its distance from the central ion falls within the interval $[ 0, d_i + \lambda ]$. For each element $i$, the value of $\lambda$ is set as half of the distance between the first and the second peak, i.e. $d_{2,i}-d_{1,i}$. For a few structures, $\lambda$ has been manually adjusted following a visual inspection of the atom positions to ensure an accurate representation of the first coordination sphere. \\
Once the number of coordinated atoms and their positions have been determined, we use the software SHAPE\cite{Casanova2004,Cirera2006} to extract the Continuous Shape Measure (CShM) \cite{Pinsky1998}. CShM provides a metric for evaluating the distance between the input geometry and a reference polyhedron model. For compounds with coordination number 4, CShM is computed with respect to tetrahedral, square, and seesaw polyhedra. For compounds with coordination number 6, the hexagonal, pentagonal pyramidal, octahedral, and trigonal prismatic polyhedra are considered. For each compound, the smallest value of the computed CShM is used to associate the complex with a reference polyhedron. For a few compounds, coordination numbers other than 4 and 6 were detected. After manually inspecting these cases, we excluded them from the present analysis.

\vspace{0.2cm}
\noindent
\textbf{Acknowledgements and Funding}\\
This project has received funding from the European Research Council (ERC) under the European Union’s Horizon 2020 research and innovation programme (grant agreement No. [948493]). Computational resources were provided by the Trinity College Research IT and the Irish Centre for High-End Computing (ICHEC). We acknowledge the discussion of results with Joris van Slageren, Biprajit Sarkar, Mauro Perfetti and Roberta Sessoli. 

\vspace{0.2cm}
\noindent
\textbf{Authors Contributions}\\
A.L. conceived the project. L.A.M. carried out the calculations and their analysis under the supervision of A.L. and the support of V.H.A.N, and V.B. L.A.M and A.L. wrote the manuscript with contributions from all the authors.

\vspace{0.2cm}
\noindent
\textbf{Data Availability}\\
The databases COMPASS\_set1, COMPASS\_set2, COMPASS\_lig1, and COMPASS\_lig2 are available at DOI: 10.5281/zenodo.13712318. 

\vspace{0.2cm}
\noindent
\textbf{Conflict of interests}\\
The authors declare that they have no competing interests.


\begin{thebibliography}{10}
\expandafter\ifx\csname url\endcsname\relax
  \def\url#1{\texttt{#1}}\fi
\expandafter\ifx\csname urlprefix\endcsname\relax\def\urlprefix{URL }\fi
\providecommand{\bibinfo}[2]{#2}
\providecommand{\eprint}[2][]{\url{#2}}

\bibitem{Lunghi2022}
\bibinfo{author}{Lunghi, A.}
\newblock \bibinfo{title}{Toward exact predictions of spin-phonon relaxation
  times: An ab initio implementation of open quantum systems theory}.
\newblock \emph{\bibinfo{journal}{Science Advances}}
  \textbf{\bibinfo{volume}{8}}, \bibinfo{pages}{eabn7880}
  (\bibinfo{year}{2022}).

\bibitem{Coey2010-qj}
\bibinfo{author}{Coey, J.}
\newblock \emph{\bibinfo{title}{Magnetism and magnetic materials}}
  (\bibinfo{publisher}{Cambridge University Press},
  \bibinfo{address}{Cambridge, England}, \bibinfo{year}{2010}).

\bibitem{ST}
\bibinfo{author}{Sucksmith, W.} \& \bibinfo{author}{Thompson, J.~E.}
\newblock \bibinfo{title}{The magnetic anisotropy of cobalt}.
\newblock \emph{\bibinfo{journal}{Proceedings of the Royal Society of London.
  Series A. Mathematical and Physical Sciences}}
  \textbf{\bibinfo{volume}{225}}, \bibinfo{pages}{362–375}
  (\bibinfo{year}{1954}).

\bibitem{Mohapatra2020}
\bibinfo{author}{Mohapatra, J.}, \bibinfo{author}{Xing, M.},
  \bibinfo{author}{Elkins, J.} \& \bibinfo{author}{Liu, J.~P.}
\newblock \bibinfo{title}{Hard and semi-hard magnetic materials based on cobalt
  and cobalt alloys}.
\newblock \emph{\bibinfo{journal}{Journal of Alloys and Compounds}}
  \textbf{\bibinfo{volume}{824}}, \bibinfo{pages}{153874}
  (\bibinfo{year}{2020}).

\bibitem{Long2018}
\bibinfo{author}{Bunting, P.~C.} \emph{et~al.}
\newblock \bibinfo{title}{A linear cobalt({II}) complex with maximal orbital
  angular momentum from a non-aufbau ground state}.
\newblock \emph{\bibinfo{journal}{Science}} \textbf{\bibinfo{volume}{362}},
  \bibinfo{pages}{eaat7319} (\bibinfo{year}{2018}).

\bibitem{Gambardella2003}
\bibinfo{author}{Gambardella, P.} \emph{et~al.}
\newblock \bibinfo{title}{Giant magnetic anisotropy of single cobalt atoms and
  nanoparticles}.
\newblock \emph{\bibinfo{journal}{Science}} \textbf{\bibinfo{volume}{300}},
  \bibinfo{pages}{1130–1133} (\bibinfo{year}{2003}).

\bibitem{Yao2016}
\bibinfo{author}{Yao, X.-N.} \emph{et~al.}
\newblock \bibinfo{title}{Two-coordinate {C}o({II}) imido complexes as
  outstanding single-molecule magnets}.
\newblock \emph{\bibinfo{journal}{Journal of the American Chemical Society}}
  \textbf{\bibinfo{volume}{139}}, \bibinfo{pages}{373–380}
  (\bibinfo{year}{2016}).

\bibitem{Vedmedenko2020}
\bibinfo{author}{Vedmedenko, E.~Y.} \emph{et~al.}
\newblock \bibinfo{title}{The 2020 magnetism roadmap}.
\newblock \emph{\bibinfo{journal}{Journal of Physics D: Applied Physics}}
  \textbf{\bibinfo{volume}{53}}, \bibinfo{pages}{453001}
  (\bibinfo{year}{2020}).

\bibitem{Reymond2015}
\bibinfo{author}{Reymond, J.-L.}
\newblock \bibinfo{title}{The chemical space project}.
\newblock \emph{\bibinfo{journal}{Accounts of Chemical Research}}
  \textbf{\bibinfo{volume}{48}}, \bibinfo{pages}{722–730}
  (\bibinfo{year}{2015}).

\bibitem{Curtarolo2013}
\bibinfo{author}{Curtarolo, S.} \emph{et~al.}
\newblock \bibinfo{title}{The high-throughput highway to computational
  materials design}.
\newblock \emph{\bibinfo{journal}{Nature Materials}}
  \textbf{\bibinfo{volume}{12}}, \bibinfo{pages}{191–201}
  (\bibinfo{year}{2013}).

\bibitem{Nandy2021}
\bibinfo{author}{Nandy, A.} \emph{et~al.}
\newblock \bibinfo{title}{Computational discovery of transition-metal
  complexes: From high-throughput screening to machine learning}.
\newblock \emph{\bibinfo{journal}{Chemical Reviews}}
  \textbf{\bibinfo{volume}{121}}, \bibinfo{pages}{9927–10000}
  (\bibinfo{year}{2021}).

\bibitem{Ale2022}
\bibinfo{author}{Lunghi, A.} \& \bibinfo{author}{Sanvito, S.}
\newblock \bibinfo{title}{Computational design of magnetic molecules and their
  environment using quantum chemistry, machine learning and multiscale
  simulations}.
\newblock \emph{\bibinfo{journal}{Nature Reviews Chemistry}}
  \textbf{\bibinfo{volume}{6}}, \bibinfo{pages}{761–781}
  (\bibinfo{year}{2022}).

\bibitem{Duan2022}
\bibinfo{author}{Duan, Y.} \emph{et~al.}
\newblock \bibinfo{title}{Data-driven design of molecular nanomagnets}.
\newblock \emph{\bibinfo{journal}{Nature Communications}}
  \textbf{\bibinfo{volume}{13}}, \bibinfo{pages}{7626} (\bibinfo{year}{2022}).

\bibitem{Janet2018}
\bibinfo{author}{Janet, J.~P.}, \bibinfo{author}{Chan, L.} \&
  \bibinfo{author}{Kulik, H.~J.}
\newblock \bibinfo{title}{Accelerating chemical discovery with machine
  learning: Simulated evolution of spin crossover complexes with an artificial
  neural network}.
\newblock \emph{\bibinfo{journal}{The Journal of Physical Chemistry Letters}}
  \textbf{\bibinfo{volume}{9}}, \bibinfo{pages}{1064–1071}
  (\bibinfo{year}{2018}).

\bibitem{Sanvito2017}
\bibinfo{author}{Sanvito, S.} \emph{et~al.}
\newblock \bibinfo{title}{Accelerated discovery of new magnets in the {H}eusler
  alloy family}.
\newblock \emph{\bibinfo{journal}{Science Advances}}
  \textbf{\bibinfo{volume}{3}}, \bibinfo{pages}{e1602241}
  (\bibinfo{year}{2017}).

\bibitem{Atanasov2015}
\bibinfo{author}{Atanasov, M.} \emph{et~al.}
\newblock \bibinfo{title}{First principles approach to the electronic
  structure, magnetic anisotropy and spin relaxation in mononuclear
  3d-transition metal single molecule magnets}.
\newblock \emph{\bibinfo{journal}{Coordination Chemistry Reviews}}
  \textbf{\bibinfo{volume}{289–290}}, \bibinfo{pages}{177–214}
  (\bibinfo{year}{2015}).

\bibitem{GomezCoca2013}
\bibinfo{author}{Gomez-Coca, S.}, \bibinfo{author}{Cremades, E.},
  \bibinfo{author}{Aliaga-Alcalde, N.} \& \bibinfo{author}{Ruiz, E.}
\newblock \bibinfo{title}{Mononuclear single-molecule magnets: Tailoring the
  magnetic anisotropy of first-row transition-metal complexes}.
\newblock \emph{\bibinfo{journal}{Journal of the American Chemical Society}}
  \textbf{\bibinfo{volume}{135}}, \bibinfo{pages}{7010–7018}
  (\bibinfo{year}{2013}).

\bibitem{Chopek2006}
\bibinfo{author}{Chłopek, K.}, \bibinfo{author}{Bothe, E.},
  \bibinfo{author}{Neese, F.}, \bibinfo{author}{Weyherm\"{u}ller, T.} \&
  \bibinfo{author}{Wieghardt, K.}
\newblock \bibinfo{title}{Molecular and electronic structures of tetrahedral
  complexes of nickel and cobalt containing {N}, {N}‘-disubstituted,
  bulkyo-diiminobenzosemiquinonate(1-) $\pi$-radical ligands}.
\newblock \emph{\bibinfo{journal}{Inorganic Chemistry}}
  \textbf{\bibinfo{volume}{45}}, \bibinfo{pages}{6298–6307}
  (\bibinfo{year}{2006}).

\bibitem{Lococciolo2024}
\bibinfo{author}{Lococciolo, G.} \emph{et~al.}
\newblock \bibinfo{title}{Oxygen-donor metalloligands induce slow magnetization
  relaxation in zero field for a cobalt({II}) complex with {CoO$_4$} motif}.
\newblock \emph{\bibinfo{journal}{Inorganic Chemistry}}
  \textbf{\bibinfo{volume}{63}}, \bibinfo{pages}{5652–5663}
  (\bibinfo{year}{2024}).

\bibitem{Gupta2023}
\bibinfo{author}{Gupta, S.~K.} \emph{et~al.}
\newblock \bibinfo{title}{Air-stable four-coordinate cobalt({II}) single-ion
  magnets: experimental and ab initio ligand field analyses of correlations
  between dihedral angles and magnetic anisotropy}.
\newblock \emph{\bibinfo{journal}{Chemical Science}}
  \textbf{\bibinfo{volume}{14}}, \bibinfo{pages}{6355–6374}
  (\bibinfo{year}{2023}).

\bibitem{Moseley2022}
\bibinfo{author}{Moseley, D.~H.} \emph{et~al.}
\newblock \bibinfo{title}{Comprehensive studies of magnetic transitions and
  spin–phonon couplings in the tetrahedral cobalt complex
  {C}o({A}s{P}h$_3$)$_2${I}$_2$}.
\newblock \emph{\bibinfo{journal}{Inorganic Chemistry}}
  \textbf{\bibinfo{volume}{61}}, \bibinfo{pages}{17123–17136}
  (\bibinfo{year}{2022}).

\bibitem{Groom2016}
\bibinfo{author}{Groom, C.~R.}, \bibinfo{author}{Bruno, I.~J.},
  \bibinfo{author}{Lightfoot, M.~P.} \& \bibinfo{author}{Ward, S.~C.}
\newblock \bibinfo{title}{The cambridge structural database}.
\newblock \emph{\bibinfo{journal}{Acta Crystallographica Section B Structural
  Science, Crystal Engineering and Materials}} \textbf{\bibinfo{volume}{72}},
  \bibinfo{pages}{171–179} (\bibinfo{year}{2016}).

\bibitem{molSimplify}
\bibinfo{author}{Ioannidis, E.~I.}, \bibinfo{author}{Gani, T. Z.~H.} \&
  \bibinfo{author}{Kulik, H.~J.}
\newblock \bibinfo{title}{mol{S}implify: A toolkit for automating discovery in
  inorganic chemistry}.
\newblock \emph{\bibinfo{journal}{Journal of Computational Chemistry}}
  \textbf{\bibinfo{volume}{37}}, \bibinfo{pages}{2106--2117}
  (\bibinfo{year}{2016}).

\bibitem{KumarSahu2023}
\bibinfo{author}{Kumar~Sahu, P.}, \bibinfo{author}{Kharel, R.},
  \bibinfo{author}{Shome, S.}, \bibinfo{author}{Goswami, S.} \&
  \bibinfo{author}{Konar, S.}
\newblock \bibinfo{title}{Understanding the unceasing evolution of {C}o({II})
  based single-ion magnets}.
\newblock \emph{\bibinfo{journal}{Coordination Chemistry Reviews}}
  \textbf{\bibinfo{volume}{475}}, \bibinfo{pages}{214871}
  (\bibinfo{year}{2023}).

\bibitem{Fataftah2014}
\bibinfo{author}{Fataftah, M.~S.}, \bibinfo{author}{Zadrozny, J.~M.},
  \bibinfo{author}{Rogers, D.~M.} \& \bibinfo{author}{Freedman, D.~E.}
\newblock \bibinfo{title}{A mononuclear transition metal single-molecule magnet
  in a nuclear spin-free ligand environment}.
\newblock \emph{\bibinfo{journal}{Inorganic Chemistry}}
  \textbf{\bibinfo{volume}{53}}, \bibinfo{pages}{10716–10721}
  (\bibinfo{year}{2014}).

\bibitem{Rechkemmer2016}
\bibinfo{author}{Rechkemmer, Y.} \emph{et~al.}
\newblock \bibinfo{title}{A four-coordinate cobalt({II}) single-ion magnet with
  coercivity and a very high energy barrier}.
\newblock \emph{\bibinfo{journal}{Nature Communications}}
  \textbf{\bibinfo{volume}{7}}, \bibinfo{pages}{10467} (\bibinfo{year}{2016}).

\bibitem{Raza2023}
\bibinfo{author}{Raza, A.} \& \bibinfo{author}{Perfetti, M.}
\newblock \bibinfo{title}{Electronic structure and magnetic anisotropy design
  of functional metal complexes}.
\newblock \emph{\bibinfo{journal}{Coordination Chemistry Reviews}}
  \textbf{\bibinfo{volume}{490}}, \bibinfo{pages}{215213}
  (\bibinfo{year}{2023}).

\bibitem{Gatteschi2006-hw}
\bibinfo{author}{Gatteschi, D.}, \bibinfo{author}{Sessoli, R.} \&
  \bibinfo{author}{Villain, J.}
\newblock \emph{\bibinfo{title}{Molecular Nanomagnets}}.
\newblock Mesoscopic Physics and Nanotechnology (\bibinfo{publisher}{Oxford
  University Press}, \bibinfo{address}{London, England}, \bibinfo{year}{2006}).

\bibitem{Sirenko2024}
\bibinfo{author}{Sirenko, V.}, \bibinfo{author}{Bartolomé~Usieto, F.} \&
  \bibinfo{author}{Bartolomé, J.}
\newblock \bibinfo{title}{The paradigm of magnetic molecule in quantum matter:
  Slow molecular spin relaxation}.
\newblock \emph{\bibinfo{journal}{Low Temperature Physics}}
  \textbf{\bibinfo{volume}{50}}, \bibinfo{pages}{431–445}
  (\bibinfo{year}{2024}).

\bibitem{Atanasov2013}
\bibinfo{author}{Atanasov, M.}, \bibinfo{author}{Zadrozny, J.~M.},
  \bibinfo{author}{Long, J.~R.} \& \bibinfo{author}{Neese, F.}
\newblock \bibinfo{title}{A theoretical analysis of chemical bonding, vibronic
  coupling, and magnetic anisotropy in linear iron({II}) complexes with
  single-molecule magnet behavior}.
\newblock \emph{\bibinfo{journal}{Chem. Sci.}} \textbf{\bibinfo{volume}{4}},
  \bibinfo{pages}{139–156} (\bibinfo{year}{2013}).

\bibitem{Zadrozny2013}
\bibinfo{author}{Zadrozny, J.~M.} \emph{et~al.}
\newblock \bibinfo{title}{Slow magnetization dynamics in a series of
  two-coordinate iron({II}) complexes}.
\newblock \emph{\bibinfo{journal}{Chem. Sci.}} \textbf{\bibinfo{volume}{4}},
  \bibinfo{pages}{125–138} (\bibinfo{year}{2013}).

\bibitem{Zadrozny2013_2}
\bibinfo{author}{Zadrozny, J.~M.} \emph{et~al.}
\newblock \bibinfo{title}{Magnetic blocking in a linear iron({I}) complex}.
\newblock \emph{\bibinfo{journal}{Nature Chemistry}}
  \textbf{\bibinfo{volume}{5}}, \bibinfo{pages}{577–581}
  (\bibinfo{year}{2013}).

\bibitem{Casanova2004}
\bibinfo{author}{Casanova, D.} \emph{et~al.}
\newblock \bibinfo{title}{Minimal distortion pathways in polyhedral
  rearrangements}.
\newblock \emph{\bibinfo{journal}{Journal of the American Chemical Society}}
  \textbf{\bibinfo{volume}{126}}, \bibinfo{pages}{1755–1763}
  (\bibinfo{year}{2004}).

\bibitem{Cirera2006}
\bibinfo{author}{Cirera, J.}, \bibinfo{author}{Ruiz, E.} \&
  \bibinfo{author}{Alvarez, S.}
\newblock \bibinfo{title}{Shape and spin state in four‐coordinate
  transition‐metal complexes: The case of the d$^6$ configuration}.
\newblock \emph{\bibinfo{journal}{Chemistry – A European Journal}}
  \textbf{\bibinfo{volume}{12}}, \bibinfo{pages}{3162–3167}
  (\bibinfo{year}{2006}).

\bibitem{Szymanski2023}
\bibinfo{author}{Szymanski, N.~J.} \emph{et~al.}
\newblock \bibinfo{title}{An autonomous laboratory for the accelerated
  synthesis of novel materials}.
\newblock \emph{\bibinfo{journal}{Nature}} \textbf{\bibinfo{volume}{624}},
  \bibinfo{pages}{86–91} (\bibinfo{year}{2023}).

\bibitem{PyzerKnapp2022}
\bibinfo{author}{Pyzer-Knapp, E.~O.} \emph{et~al.}
\newblock \bibinfo{title}{Accelerating materials discovery using artificial
  intelligence, high performance computing and robotics}.
\newblock \emph{\bibinfo{journal}{npj Computational Materials}}
  \textbf{\bibinfo{volume}{8}}, \bibinfo{pages}{84} (\bibinfo{year}{2022}).

\bibitem{Oganov2019}
\bibinfo{author}{Oganov, A.~R.}, \bibinfo{author}{Pickard, C.~J.},
  \bibinfo{author}{Zhu, Q.} \& \bibinfo{author}{Needs, R.~J.}
\newblock \bibinfo{title}{Structure prediction drives materials discovery}.
\newblock \emph{\bibinfo{journal}{Nature Reviews Materials}}
  \textbf{\bibinfo{volume}{4}}, \bibinfo{pages}{331–348}
  (\bibinfo{year}{2019}).

\bibitem{Zahradnkov2023}
\bibinfo{author}{Zahradníková, E.}, \bibinfo{author}{Sutter, J.-P.},
  \bibinfo{author}{Halaš, P.} \& \bibinfo{author}{Drahoš, B.}
\newblock \bibinfo{title}{Trigonal prismatic coordination geometry imparted by
  a macrocyclic ligand: an approach to large axial magnetic anisotropy for
  {C}o({II})}.
\newblock \emph{\bibinfo{journal}{Dalton Transactions}}
  \textbf{\bibinfo{volume}{52}}, \bibinfo{pages}{18513–18524}
  (\bibinfo{year}{2023}).

\bibitem{Brachakov2020}
\bibinfo{author}{Brachňaková, B.} \emph{et~al.}
\newblock \bibinfo{title}{Stereochemistry of coordination polyhedra vs. single
  ion magnetism in penta- and hexacoordinated {C}o({II}) complexes with
  tridentate rigid ligands}.
\newblock \emph{\bibinfo{journal}{Dalton Transactions}}
  \textbf{\bibinfo{volume}{49}}, \bibinfo{pages}{1249–1264}
  (\bibinfo{year}{2020}).

\bibitem{Canaj2019}
\bibinfo{author}{Canaj, A.~B.} \emph{et~al.}
\newblock \bibinfo{title}{Insight into {D}$_{6h}$ symmetry: Targeting strong
  axiality in stable dysprosium({III}) hexagonal bipyramidal single‐ion
  magnets}.
\newblock \emph{\bibinfo{journal}{Angewandte Chemie International Edition}}
  \textbf{\bibinfo{volume}{58}}, \bibinfo{pages}{14146–14151}
  (\bibinfo{year}{2019}).

\bibitem{Li2019}
\bibinfo{author}{Li, Z.}, \bibinfo{author}{Zhai, Y.}, \bibinfo{author}{Chen,
  W.}, \bibinfo{author}{Ding, Y.} \& \bibinfo{author}{Zheng, Y.}
\newblock \bibinfo{title}{Air‐stable hexagonal bipyramidal dysprosium({III})
  single‐ion magnets with nearly perfect d6h local symmetry}.
\newblock \emph{\bibinfo{journal}{Chemistry – A European Journal}}
  \textbf{\bibinfo{volume}{25}}, \bibinfo{pages}{16219–16224}
  (\bibinfo{year}{2019}).

\bibitem{Zhu2021}
\bibinfo{author}{Zhu, Z.} \emph{et~al.}
\newblock \bibinfo{title}{Air-stable chiral single-molecule magnets with record
  anisotropy barrier exceeding 1800 k}.
\newblock \emph{\bibinfo{journal}{Journal of the American Chemical Society}}
  \textbf{\bibinfo{volume}{143}}, \bibinfo{pages}{10077–10082}
  (\bibinfo{year}{2021}).

\bibitem{Atomsk}
\bibinfo{author}{Hirel, P.}
\newblock \bibinfo{title}{Atomsk: A tool for manipulating and converting atomic
  data file}.
\newblock \emph{\bibinfo{journal}{Computer Physics Communications}}
  \textbf{\bibinfo{volume}{197}}, \bibinfo{pages}{212–219}
  (\bibinfo{year}{2015}).

\bibitem{Bartk2013}
\bibinfo{author}{Bartók, A.~P.}, \bibinfo{author}{Kondor, R.} \&
  \bibinfo{author}{Csányi, G.}
\newblock \bibinfo{title}{On representing chemical environments}.
\newblock \emph{\bibinfo{journal}{Physical Review B}}
  \textbf{\bibinfo{volume}{87}}, \bibinfo{pages}{184115}
  (\bibinfo{year}{2013}).

\bibitem{Bannwarth2020}
\bibinfo{author}{Bannwarth, C.} \emph{et~al.}
\newblock \bibinfo{title}{Extended tight‐binding quantum chemistry methods}.
\newblock \emph{\bibinfo{journal}{WIREs Computational Molecular Science}}
  \textbf{\bibinfo{volume}{11}}, \bibinfo{pages}{e1493} (\bibinfo{year}{2020}).

\bibitem{GFN2}
\bibinfo{author}{Bannwarth, C.}, \bibinfo{author}{Ehlert, S.} \&
  \bibinfo{author}{Grimme, S.}
\newblock \bibinfo{title}{{GFN}2-x{TB}—an accurate and broadly parametrized
  self-consistent tight-binding quantum chemical method with multipole
  electrostatics and density-dependent dispersion contributions}.
\newblock \emph{\bibinfo{journal}{Journal of Chemical Theory and Computation}}
  \textbf{\bibinfo{volume}{15}}, \bibinfo{pages}{1652--1671}
  (\bibinfo{year}{2019}).

\bibitem{neese2020orca}
\bibinfo{author}{Neese, F.}, \bibinfo{author}{Wennmohs, F.},
  \bibinfo{author}{Becker, U.} \& \bibinfo{author}{Riplinger, C.}
\newblock \bibinfo{title}{The orca quantum chemistry program package}.
\newblock \emph{\bibinfo{journal}{The Journal of chemical physics}}
  \textbf{\bibinfo{volume}{152}}, \bibinfo{pages}{224108}
  (\bibinfo{year}{2020}).

\bibitem{Becke1988}
\bibinfo{author}{Becke, A.~D.}
\newblock \bibinfo{title}{Density-functional exchange-energy approximation with
  correct asymptotic behavior}.
\newblock \emph{\bibinfo{journal}{Physical Review A}}
  \textbf{\bibinfo{volume}{38}}, \bibinfo{pages}{3098–3100}
  (\bibinfo{year}{1988}).

\bibitem{Perdew1986}
\bibinfo{author}{Perdew, J.~P.}
\newblock \bibinfo{title}{Density-functional approximation for the correlation
  energy of the inhomogeneous electron gas}.
\newblock \emph{\bibinfo{journal}{Physical Review B}}
  \textbf{\bibinfo{volume}{33}}, \bibinfo{pages}{8822–8824}
  (\bibinfo{year}{1986}).

\bibitem{Grimme2010}
\bibinfo{author}{Grimme, S.}, \bibinfo{author}{Antony, J.},
  \bibinfo{author}{Ehrlich, S.} \& \bibinfo{author}{Krieg, H.}
\newblock \bibinfo{title}{A consistent and accurateab initioparametrization of
  density functional dispersion correction ({DFT}-{D}) for the 94 elements
  {H}-{P}u}.
\newblock \emph{\bibinfo{journal}{The Journal of Chemical Physics}}
  \textbf{\bibinfo{volume}{132}}, \bibinfo{pages}{154104}
  (\bibinfo{year}{2010}).

\bibitem{atanasov2011}
\bibinfo{author}{Atanasov, M.}, \bibinfo{author}{Ganyushin, D.},
  \bibinfo{author}{Sivalingam, K.} \& \bibinfo{author}{Neese, F.}
\newblock \bibinfo{title}{Molecular electronic structures of transition metal
  complexes ii}.
\newblock In \bibinfo{editor}{Mingos, D. M.~P.}, \bibinfo{editor}{Day, P.} \&
  \bibinfo{editor}{Dahl, J.~P.} (eds.) \emph{\bibinfo{booktitle}{Structure and
  Bonding}}, vol. \bibinfo{volume}{143}, \bibinfo{pages}{149--220}
  (\bibinfo{publisher}{Springer Berlin Heidelberg}, \bibinfo{year}{2011}).

\bibitem{Lang2020}
\bibinfo{author}{Lang, L.}, \bibinfo{author}{Atanasov, M.} \&
  \bibinfo{author}{Neese, F.}
\newblock \bibinfo{title}{Improvement of ab initio ligand field theory by means
  of multistate perturbation theory}.
\newblock \emph{\bibinfo{journal}{The Journal of Physical Chemistry A}}
  \textbf{\bibinfo{volume}{124}}, \bibinfo{pages}{1025–1037}
  (\bibinfo{year}{2020}).

\bibitem{Pinsky1998}
\bibinfo{author}{Pinsky, M.} \& \bibinfo{author}{Avnir, D.}
\newblock \bibinfo{title}{Continuous symmetry measures. 5. the classical
  polyhedra}.
\newblock \emph{\bibinfo{journal}{Inorganic Chemistry}}
  \textbf{\bibinfo{volume}{37}}, \bibinfo{pages}{5575–5582}
  (\bibinfo{year}{1998}).

\end{thebibliography}
\end{document}